\documentclass[11pt,a4paper]{article}
\usepackage{jheppub_kim}
\usepackage{url}
\usepackage[utf8]{inputenc}
\usepackage{subfigure}
\usepackage{pdflscape}
\usepackage{amsmath}
\usepackage{amssymb}
\usepackage{dcolumn}
\usepackage{bm}
\usepackage{color}
\usepackage{epsfig}
\usepackage{amsfonts}
\usepackage{graphicx}
\usepackage{subfigure}
\usepackage{dcolumn}
\usepackage{amssymb, amsmath, amsfonts, amsthm, graphicx}
\usepackage{pdflscape}
\def\bea{\begin{eqnarray}}
\def\eea{\end{eqnarray}}

\def\be{\begin{equation}}
\def\ee{\end{equation}}

\def\udot{\dot{u}}

\def\ex{e_1{}^1}
\def\ey{e_2{}^2}

\def\y{\vartheta}
\def\z{\varphi}

\PassOptionsToPackage{linktocpage}{hyperref}
\def\case#1/#2{\textstyle\frac{#1}{#2}}

\def\udot{\dot{u}}
\def\ex{e_1{}^1}
\def\ey{e_2{}^2}

\def\y{\vartheta}
\def\z{\varphi}

\def\be{\begin{equation}}
\def\ee{\end{equation}}
\def\bea{\begin{eqnarray}}
\def\eea{\end{eqnarray}}

\begin{document}

\title{Kantowski-Sachs Einstein-\ae ther perfect fluid models}

\author[a]{Joey Latta}
\author[b]{Genly Leon}
\author[c]{Andronikos Paliathanasis}

\affiliation[a]{Department of Mathematics and Statistics, Dalhousie University, Halifax, Nova Scotia, Canada  B3H 3J5}
\affiliation[b]{Instituto de F\'{\i}sica, Pontificia Universidad de Cat\'olica de Valpara\'{\i}so, Casilla 4950, Valpara\'{\i}so, Chile}
\affiliation[c]{Instituto de Ciencias F\'{\i}sicas y Matem\'{a}ticas, Universidad Austral de Chile, Valdivia, Chile}

\emailAdd{lattaj@mathstat.dal.ca}

\emailAdd{genly.leon@pucv.cl}

\emailAdd{anpaliat@phys.uoa.gr}

\abstract{We investigate Kantowski-Sachs models in Einstein-\ae ther theory with a perfect fluid source using the singularity analysis to prove the integrability of the field equations and dynamical system tools to study the evolution. We find an inflationary source at early times, and an inflationary sink at late times, for a wide region in the parameter space. The results by A. A. Coley, G. Leon, P. Sandin and J. Latta (JCAP {\bf 12}, 010, 2015), are then re-obtained as particular cases. Additionally, we select other values for the non-GR parameters which are consistent with current constraints, getting a very rich phenomenology. In particular, we find solutions with infinite shear, zero curvature, and infinite matter energy density in comparison with the Hubble scalar. We also have stiff-like future attractors, anisotropic late-time attractors, or both, in some special cases. Such results are developed analytically, and then verified by numerics. Finally, the physical interpretation of the new critical points is discussed.}

\keywords{Kantowski-Sachs, Einstein-\ae ther theory, perfect fluid}
\arxivnumber{1606.08586}
\maketitle

\section{Introduction}

The cosmological acceleration of the Universe is a challenge to our knowledge of physics, since it cannot be
described within the framework of general relativity (GR) for a matter content satisfying the strong energy condition.
Thus, in order to explain it one should either keep GR and
modify the matter content of the universe, introducing the concept of dark energy
\cite{Copeland:2006wr,Cai:2009zp}, or modify the gravitational sector itself. In
particular, one can modify gravity by constructing various extensions of the
Einstein-Hilbert action, such as $f(R)$ gravity \cite{DeFelice:2010aj}
$f(G)$ gravity \cite{Nojiri:2005jg}, Lovelock
gravity \cite{Lovelock:1971yv},   Ho\v{r}ava-Lifshitz gravity \cite{Horava:2009uw},
massive gravity \cite{deRham:2010kj}, galileon modifications \cite{Nicolis:2008in}, etc.
(for reviews see  \cite{modgravity}).

One interesting class of gravitational modification is the Einstein-aether theories
(\AE-theories), which were investigated systematically in the last fifteen years
\cite{Jacobson:2000xp,Eling:2004dk,Jacobson,DJ,Carruthers:2010ii,Kanno:2006ty,
Zlosnik:2006zu,
Jacobson:2010mx,Foster:2005dk,Eling:2006ec,Eling:2006df,Yagi:2013ava,Foster:2006az,
Eling:2005zq,
Garfinkle:2007bk,Berglund:2012fk,Foster:2007gr,Eling:2007xh,Konoplya:2006rv,
Heinicke:2005bp,Foster:2005ec,Foster:2005fr,ArmendarizPicon:2010rs,Jacobson:2013xta,
Bonvin:2007ap,Jacobson:2007fh,Dai:2008sf,Konoplya:2006ar,Haghani:2014ita,Meng:2011wc,
Withers:2009qg,Meng:2012zza, Jacobson:2014mda,
Balakin:2014tza,Nakashima:2011fu,Xie:2008ry,Gao:2013im,Nakashima:2010nq,Gurses:2009zz,
Garfinkle:2012rr,Gurses:2015bjt,Pasqua:2015rxa, Xie:2008ooa},
although the work on the subject of \AE-theories started a long ago in special cases
\cite{
Gasperini:1986ym,Kostelecky:1989jp}. Recently, in \cite{Coley:2015qqa}, two of us studied spherically
symmetric \AE-theories with a  perfect fluid matter source in a comprehensive manner. We have derived a well-posed system of first order partial differential
evolution equations in two variables with restrictions. Introducing normalized variables, we obtained a set of equations well suited for numerical computations and
for the study of the qualitative properties of the models.
\AE-theories consist of GR  coupled, at second derivative order, to a dynamical timelike unit vector field, the aether, $u^a$. This vector can be thought of as the 4-velocity of a preferred frame.

The energy momentum tensor of the \ae ther is built by contraction of
$\nabla_{a}u^{c}\nabla_{b}u^{d}
$ with the tensor
\begin{equation}
K^{a b}{}_{c d}\equiv c_{1}g^{a b}g_{c d} + c_{2}
\delta_{c}^{a}\delta_{d}^{b} + c_{3}\delta_{d}^{a}
\delta_{c}^{b} + c_{4}u^{a}u^{b}g_{c d},
\end{equation}
that depends on four constants.
 We can study different models with different dimensionless parameters $c_i$. It is
convenient to
make a reparameterization of the aether
parameters, analogous to the one given in~\cite{Jacobson:2013xta}:
\begin{displaymath}
c_\theta = c_2 + (c_1 + c_3)/3,\ c_\sigma = c_1 + c_3,\ c_\omega = c_1 - c_3,\ c_a = c_4 -
c_1,
\end{displaymath}
where the new parameters correspond to terms in the Lagrangian relating to expansion,
shear, acceleration, and twist of the aether respectively.  As we know,  spherically
symmetric models are hypersurface orthogonal, thus, the \ae ther field has vanishing twist and therefore it is independent of the twist parameter $c_\omega$ \cite{Jacobson:2013xta}. This freedom in the parameters choice can be used to set $c_4=0$  \cite{Jacobson}.
 A second condition on the $c_i$ can effectively be specified  by a renormalization of
the Newtonian gravitational constant $G$. From \cite{Jacobson} we obtain that $G_N =
G\left(1 -  \frac{1}{2}(c_\sigma+c_\omega+c_a)\right)^{-1}$. As long as $(c_\sigma+c_\omega+c_a) <2$, so that
 the gravitational constant is positive, we can effectively renormalize and specify $c_\sigma+c_\omega+c_a$.
The remaining two non-trivial constant parameters in the model must satisfy additional constraints.

For the above models the values of the $c_i, i=1 \ldots 4$ must be consistent with all
observations. In general, if the magnitudes of all   the $c_i$ are
non-zero and small, say, less than $10^{-2}$, then the models will be physical
 \cite{Jacobson,Barausse:2011pu}. There are also a number of self-consistency requirements
\cite{Jacobson,Barausse:2011pu}.  If we study the models in the early universe, where the constants $c_i$
can be replaced with evolving parameters \cite{Kanno:2006ty}, then the observational constraints introduced in \cite{Jacobson,Barausse:2011pu} need not be applied.

On the other hand, there is some interest in cosmological models with positive spatial
curvature (closed models) \cite{Coley:2000yc}, but they have attracted less attention since they are more
complicated mathematically. Closed Friedmann-Lemaître-Robertson-Walker (FLRW) models were investigated, e.g., in
\cite{Halliwell:1986ja,vandenHoogen:1999qq,WE,Coley:2003mj}; Kantowski-Sachs models in
\cite{Coley:2003mj,Heinzle:2004sr}; and Bianchi type IX in
\cite{WE,Heinzle:2004sr,Kitada:1991ih,vandenHoogen:1998cc} using dynamical system
techniques, while a compact phase space analysis was performed in
\cite{Coley:2000yc,Goliath:1998na,UM}.

In our research we are interested in Kantowski-Sachs models, which are defined locally as admitting a four-parameter continuous isometry group which
acts on spacelike hypersurfaces, and which possesses a three-parameter subgroup whose
orbits are 2-surfaces of constant curvature. This implies that models possess spherical
symmetry, combined with a translational symmetry \cite{WE,Coley:2003mj,Collins:1977fg}.
Earlier references to Kantowski-Sachs models can be found in \cite{Collins:1977fg}. The
Kantowski-Sachs models can be obtained from the locally rotationally symmetric
Bianchi type IX models by a Lie contraction, and thus, they can appear as invariant sets
in the boundary of the phase space of locally rotationally symmetric Bianchi IX non-tilted
perfect fluid models \cite{WE,UM}.

The recollapse problem was solved in  \cite{Collins:1977fg} for all
general relativistic Kantowski-Sachs models in which the matter content is a perfect fluid satisfying
reasonable energy conditions. These models exhibit, in almost all cases, past asymptotes to a
Big-Bang singularity and future asymptotes to a Big Crunch \cite{WE}. Classically, these past and future singularities can be an anisotropic structure such as a barrel, cigar or a pancake, or an isotropic point--like structure depending on the initial conditions on anisotropic shear and matter \cite{Collins:1977fg}. Furthermore,
in \cite{Coley:2000yc}  closed FLRW models were investigated containing a perfect fluid and an exponential scalar field,
while in \cite{Heinzle:2004sr} both locally rotationally symmetric Bianchi type IX and
Kantowski-Sachs orthogonal perfect fluid models were investigated.
Kantowski-Sachs models were also investigated in some cosmological settings such that modified
gravity and scalar-field models
\cite{Leon:2010pu,Leon:2014dea,Leon:2013bra,Fadragas:2013ina}. Closed FLRW
models were also investigated in \cite{Leon:2009ce,Leon:2015via}. The global
asymptotic behaviour of closed FLRW models is that they either expand from an initial singularity, reach a maximum expansion
and thereafter recollapse to a final singularity, or else they expand forever towards a
flat power-law inflationary solution, as occurs in the Bianchi type IX
models. Now, for the Kantowski-Sachs model there are two asymptotic scenarios: (i) by qualitative analysis
it was found that all models expand from a singularity, reach a point of maximum
expansion, and then recollapse to a singularity; and (ii), it was numerically proven that
there are solutions that expand from singularities to infinitely dispersed isotropic
states and solutions that contract from infinitely dispersed isotropic states to
singularities \cite{Coley:2003mj,Heinzle:2004sr}.
Now, in our scenario, i.e. a perfect fluid in Kantowski-Sachs \AE-theory without scalar field, we also found solutions that either expand from or contract to anisotropic states. This result, to our knowledge, is new (a partial proof of this was given by two us in \cite{Coley:2015qqa}) and does not arise in GR. These solutions are a non-trivial consequence of the presence of a non-zero
Lorentz-violating vector field.

 In this paper,  we will investigate Kantowski-Sachs models in \AE-theory with a perfect fluid source, extending the results presented in  \cite{Coley:2015qqa}.
Since models of this kind may recollapse, the expansion parameter $\theta$, which is related to the Hubble parameter, is
zero at the time of maximal expansion, and thus, $\theta$-normalization does not
lead to a complete description of the dynamics. We derive the evolution equations in
terms of normalized variables, rather than $\theta$-normalized, which reduce to a
dynamical system. Our results extend and complement those found in \cite{Coley:2015qqa}. The
formalism adopted here is appropriate for the study of the qualitative properties of
astrophysical and cosmological models with values for the parameters $c_a, c_\theta, c_\sigma$ of the \ae ther field,  which are
consistent with current constraints. In particular, for the range of parameters $0\leq
\gamma < 2$, $c_\theta <  -\frac{1}{3}$ and $c_\sigma < \frac{1}{2}$, where $\gamma$ is
the barotropic index of the perfect fluid, we find an inflationary source at early times,
and an inflationary sink at late times. In the special case
$c_\sigma=\frac{1}{2}(1- c^2)  \geq 0, c_a =  -\frac{d}{(1+d)}
c_\sigma \leq 0, c_\theta=0$, and under variable rescalings the system (5.27) investigated in \cite{Coley:2015qqa} is recovered and the
results presented there are re-obtained. Additionally, we discuss two new cases that arise for a special
selection of the free parameters in accordance to physical bounds (as discussed in
\cite{Jacobson,Barausse:2011pu,Kanno:2006ty} and summarized in Appendix A of \cite{Coley:2015qqa}). For example, assuming
$c_a = -\frac{(c_1{^2} +  c_3{^2})}{c_1} \leq 0
 , \quad
 0 \leq c_\sigma  = {c_1} + {c_3} \leq 1,
 \quad
 c_\theta = -\frac{(c_1{^2} -  c_3{^2})}{3c_1} \leq 0,$
particularly, for the choices  $c_1<0, \frac{1}{2} (1-2 c_1)<c_3\leq
   1-c_1$  or
 $\frac{1}{4}<c_1\leq \frac{1}{2},
   \frac{1}{2} (1-2 c_1)<c_3\leq c_1$  or
	 $c_1>\frac{1}{2}, \frac{1}{2} (1-2 c_1)<c_3\leq
   1-c_1$,
we show that the phase space becomes unbounded. For these ranges of
parameters, we demonstrate the existence of solutions with infinite shear, zero curvature and infinite
mass energy density in comparison with the Hubble scalar. We also have stiff-like future
attractors, anisotropic late-time  attractors, or both, in some special cases.
Finally, in the case $c_\sigma=\frac{1}{2}(1- c^2)  \geq 0,
 c_\theta = -\frac{1}{3}(1- c^2) \leq 0, c_a = 0$, and under a time
rescaling, we show that the late-time attractors are stiff-like solutions. 		
Such results are developed analytically, and then verified numerically.

As far as concerns the critical points that are introduced by the \AE-theory, we discuss the behavior of the solutions at such points and we find conditions for the non existence of a Big Bang or a Big Crunch singularity, and compare them with that of the stability analysis of the critical point.

\section{The model}

The action of \AE-gravity reads \cite{Jacobson,Carroll:2004ai}:
\begin{equation}
S=\int d^{4}x\sqrt{-g}\left[  \frac{1}{2}R - K^{a b}{}_{c d
}\nabla_{a}u^{c}\nabla_{b}u^{d} + \lambda\left(  u^{c}u_{c
} + 1\right) + \mathcal{{L}}_m  \right]  ,\label{action}
\end{equation}
where
\begin{equation}
K^{a b}{}_{c d}\equiv c_{1}g^{a b}g_{c d} + c_{2}
\delta_{c}^{a}\delta_{d}^{b} + c_{3}\delta_{d}^{a}
\delta_{c}^{b} + c_{4}u^{a}u^{b}g_{c d}.
\end{equation}
The action (\ref{action}) contains the Einstein-Hilbert term, a kinetic term
for the \ae ther with four dimensionless coefficients $c_{i}$, and the matter-energy density $\mathcal{{L}}_m$. The constraint on the \ae ther to be time-like is guaranteed by means of the Lagrange multiplier $\lambda$ \cite{Garfinkle:2011iw}. The Lagrange multiplier can be defined by \cite{Coley:2015qqa}:
\begin{equation}
\label{definition:lambda}
\lambda = - u^b \nabla_a J^a_b-c_4 \udot_a \udot^a
\end{equation} where ${{J^a}_m}=-{{K^{ab}}_{mn}}{\nabla_b}{u^n}$, ${\dot u_a} = {u^b}{\nabla _b}{u_a}$. Furthermore, the \ae ther vector must satisfy the restrictions \cite{Coley:2015qqa}:
\begin{equation}
\label{restriction_aether}
0 = h^{b c}\nabla_a J^a_b + c_4 h^{b c} \udot_a \nabla_b u^a,
\end{equation} where $h^{b c} := g^{b c}+u^{b} u^{c}$ denotes the induced metric. \\
The convention used in this paper for metric signature is $({-}{+}{+}{+})$ and the units are chosen so that the speed of
light defined by the metric $g_{ab}$ is unity and $\kappa^2\equiv 8\pi G=1.$
The field equations from varying (\ref{action}) with respect to
$g^{ab}$ are \cite{Garfinkle:2007bk}:
\bea
{G_{ab}} &=& T^{\ae}_{ab}+T^{m}_{ab}
\label{EFE2}
 \eea
where $G_{ab}$ is the Einstein tensor of the metric $g_{ab}$. The effective \ae ther energy-momentum tensor is given by
\begin{align}
{T^{\ae}_{ab}} &= 2c_{1}(\nabla_{a}u^{c}\nabla_{b}u_{c}-
\nabla^{c}u_{a}\nabla_{c}u_{b})- 2[\nabla_{c}(u_{(a} J^{c}{}_{b)}) + \nabla_{c}(u^{c
}J_{(a b)}) - \nabla_{c}(u_{(a}J_{b)}{}^{c})] \nonumber\\
&   -2 c_4 \udot_a \udot_b   + 2\lambda u_a u_b + g_{a b}\mathcal{L}_{u}, \label{aestress}
\end{align}
 and
\begin{equation}\label{aeLagrangian}
\mathcal{L}_{u} \equiv -K^{a b}{}_{c d}\nabla_{a}u^{c
}\nabla_{b}u^{d},
\end{equation} is the \AE- lagrangian \cite{Jacobson}.
The energy momentum-tensor for the matter field is
  \be
 {T^{m}_{ab}}\equiv -2\frac{\delta \mathcal{L}_m}{\delta g^{a b}}+\mathcal{L}_m g_{a b}=
\mu
u_a u_b + p ( g_{ab} + u_a u_b).
 \ee We choose a linear equation of state for
the perfect fluid:
\be
 \label{linear_eos}
    p = (\gamma-1) \mu,
\ee
where $\gamma$ is a constant satisfying $0 \leq \gamma < 2$.

 From before, it is convenient to introduce the redefinition of constants:
 \begin{displaymath}
 c_\theta = c_2 + (c_1 + c_3)/3,\ c_\sigma = c_1 + c_3,\ c_\omega = c_1 - c_3,\ c_a = c_4
- c_1.
  \end{displaymath}

Using the Kantowski-Sachs metric \cite{kramer}:
 \be
         ds^2 = - N(t)^2 dt^2 + (\ex(t))^{-2} dx^2
                 + (\ey(t))^{-2} (d\y^2 + \sin^2 \y  d\z^2),
 \ee
and setting the lapse function to $N=1$, the Lagrangian \eqref{aeLagrangian} becomes $\mathcal{L}_{u} = - \left(c_\theta \theta^2 + 6 c_\sigma \sigma^2\right),$
and the \ae ther components reduce to
$(\mu,\ p,\ q_1,\ \pi_+) = (- c_\theta \theta^2 - 6 c_\sigma\sigma^2,\ c_\theta (2 \partial_t + \theta)\theta - 6 c_\sigma\sigma^2,
0,\ 2 c_\sigma (\partial_t + \theta)\sigma)$. While the evolution equations are given by the following algebraic-differential system \cite{Coley:2015qqa}:
\begin{subequations}
\label{system_1}
\begin{align}
&\dot{\ex}=-\frac{1}{3}\left( \theta -6\sigma \right) \ex  \label{ll.01}
\\
&\dot{\theta}=-\frac{\theta ^{2}}{3}+6\frac{C_{1}}{C_{2}}\sigma ^{2}+\frac{%
\left( 2-3\gamma \right) }{2C_{2}}\mu  \label{ll.02} \\
& \dot{\sigma}=-\frac{1}{9}\frac{C_{2}}{C_{1}}\theta ^{2}-\theta \sigma
-\sigma ^{2}+\frac{1}{3C_{1}}\mu  \label{ll.03} \\
& \dot{\mu}=-\gamma \theta \mu  \label{ll.04} \\
& \dot{K}=-\frac{2}{3}\left( \theta +3\sigma \right) K  \label{ll.05}
\end{align}%
\end{subequations}
where the algebraic equation is
\begin{equation}
K=\mu -3C_{1}\sigma ^{2}-\frac{C_{2}}{3}\theta ^{2},  \label{ll.06}
\end{equation}%
where $C_{1}=2c_{\sigma
}-1~,~C_{2}=3c_{\theta }+1.$ The choice $C_1=-1, C_2=1$
corresponds to GR.

 It is possible to write \eqref{system_1} as a system of second-order
differential equations but with fewer independent variables. For the latter
systems we apply the ARS algorithm (Ablowitz, Ramani and Segur) \cite%
{Abl1,Abl2,Abl3}. Because of the algebraic equation (\ref{ll.06}) the dynamical system
reduces to that of four first-order differential equations, (\ref{ll.01})-(%
\ref{ll.04}). Furthermore from (\ref{ll.01}) and (\ref{ll.04}) we find that
\begin{equation}
\theta=-\frac{1}{\gamma}\frac{\dot{\mu}}{\mu} \quad \mbox{\rm and} \quad \sigma=\frac{\theta}{6}+%
\frac{1}{2}\frac{\dot\ex}{\ex}  \label{ll.07a}
\end{equation}
from which, if we substitute into (\ref{ll.02}) and (\ref{ll.03}), we have the
following system of second-order differential equations with respect to the
variables $\mu$ and $\ex$. \ The system is
\begin{subequations}
\label{2.14}
\begin{align}
0 & =9C_{1}\gamma^{2}\mu^{2}\left( \dot\ex\right) ^{2}-6C_{1}\gamma
\ex\mu\dot\ex\dot{\mu}+  \notag \\
& +\left(\ex\right)^{2}\left( 3\left( 2-3\gamma\right) \gamma^{2}\mu^{3}+\left(
C_{1}-2C_{2}-6C_{2}\gamma\right) \dot{\mu}^{2}+6C_{2}\gamma\mu\ddot{\mu }%
\right)  \label{ll.08} \\
0 & =9C_{1}\left( C_{2}-C_{1}\right) \gamma^{2}\mu^{2}\left( \dot\ex\right) ^{2}+3\gamma^{2}\left(\ex\right)^{2}\left( 4C_{2}+3C_{1}\gamma
-2C_{1}\right) \mu^{3}+  \notag \\
& -\left(\ex\right)^{2}\left( C_{1}+C_{2}\right) \left( C_{1}+4C_{2}\right) \dot {\mu}%
^{2}+6C_{1}\gamma \ex\mu\left( \left( C_{1}+4C_{2}\right) \dot {e}\dot{\mu}%
-4C_{2}\gamma\mu\ddot\ex\right)  \label{ll.09}
\end{align}

With the method of singularity analysis we prove the integrability of
the system \eqref{2.14}.

\section{Integrability of the field equations}
A dynamical system can either be studied by various approaches or
be solved numerically. However, in order to prove that there exists an actual
solution the integrability of the system should be studied. The integrability in
gravitational theories is a subject of special interest. For the silent
universe the integrability of the irrotational models has been proven in \cite{silent1}, while in case of the Szekeres system the integrability has been analyzed with the method of Darboux
polynomials in \cite{silent2}. In this section we will use the method of movable singularities in order to study the integrability of the field equations. Singularity analysis has been applied previously to various cosmological models \cite{miritzis,helmi,cots} and most recently in the modified
theories of gravity namely $f\left( R\right) $ and $f\left( T\right) $ \cite%
{sinFR,sinFT}.
In order to perform the singularity analysis we follow the ARS (Ablowitz, Ramani and Segur) \cite{Abl1,Abl2,Abl3}
algorithm. Specifically the steps that we follow are: (a) determine
dominant solution, (b) find the resonances and (c) write the solution in a
Laurent Series and prove the consistency of the solution.

We substitute $\left( \mu \left( t\right) ,\ex\left( t\right) \right)
=\left( \mu _{0}\tau ^{p},e_{0}\tau ^{q}\right) ~$,~($\tau =t-t_{0}$) into (%
\ref{ll.08}) and (\ref{ll.09}) and we determine the dominant behavior. Easily
we find that
\end{subequations}
\begin{equation}
p=-2~,~q=-\frac{2}{3\gamma }~,~\mu _{0}=\frac{4C_{2}}{3\gamma ^{2}}
\label{ll.10}
\end{equation}%
\begin{equation}
p=-2,~q=2\left( 1-\frac{1}{\gamma }\right) ~,~\mu _{0}=\frac{4}{3\gamma ^{2}}%
\left( \left( 3\gamma -2\right) C_{1}+C_{2}\right).  \label{ll.11}
\end{equation}

We observe that (\ref{ll.10}) and (\ref{ll.11}) are also
solutions of the field equations \eqref{2.14}. This is a particular solution and holds for specific initial conditions because the free
parameter is only the position of the singularity, $t_{0}$. It is important
to note here that solution (\ref{ll.10}) is that which is given by the
application of the zeroth-order invariants of the Lie symmetry vector $t\partial
_{t}-2\mu \partial _{\mu }-\frac{2}{3\gamma }\ex\partial _{\ex}~$of the
system \eqref{2.14}.

We continue with the determination of the resonances for the dominant terms (%
\ref{ll.10}). In order to do that we substitute into the system \eqref{2.14},
\begin{equation}
\mu \left( t\right) =\frac{4}{3}\frac{C_{2}}{\gamma ^{2}}%
t^{-2}+mt^{-2+s}~,~\ex\left( t\right) =e_{0}t^{-\frac{2}{3\gamma }%
}+nt^{-2+s},  \label{ll.13}
\end{equation}%
where $m,~n$ are two arbitrary parameters. We linearize
around $m=0$ and $n=0$, i.e. $m^{2}\rightarrow 0,~n^{2}\rightarrow 0$,~$%
mn\rightarrow 0$ etc.  We obtain two linear algebraic equations on $%
m$ and $n$. The system should have arbitrary solution \ for any values of $m$
and $n,~$that is, the determinant of the matrix which defines the linear
system has to vanish. From the latter we find the algebraic equation
\begin{equation}
\left( 1+s\right) \left( 8-9\gamma +3\gamma s\right) \left( 2-6\gamma
+3\gamma s\right) \left( 4-6\gamma +3\gamma s\right) =0  \label{ll.14}
\end{equation}%
in which the solution with respect to $s$ gives.
\begin{equation}
s_{1}=-1~,~s_{2}=\frac{2}{3\gamma }\left( 3\gamma -2\right) ~,~s_{3}=\frac{2%
}{3}\left( 3\gamma -1\right) ~,~s_{4}=\frac{\left( 9\gamma -8\right) }{%
3\gamma },  \label{ll.15}
\end{equation}%
The resonance $s_{1}=-1$, is important for the existence of the singularity
while it provides us with the information that the singularity analysis have
been done correctly. The other three resonances give the positions of the
three integration constants in the series expansion.  Recall that the fourth constant of integration is
the position of the singularity, $t_{0}$. An important observation here is
that for $\gamma \in \lbrack 1,2)$ all the resonances (\ref{ll.15}) are
positive meaning that the solution is given by a Right Painlev\'{e} Series.

In order for the equations to pass the singularity analysis we have to check the
consistency of the solution. For that reason we select $\gamma =1$, a dust
fluid, from which we have that the dominant solution $\mu _{d}\left(
t\right) =\frac{4}{3}C_{2}t^{-2}~,~\ex_{d}\left( t\right) =e_{0}t^{-\frac{2}{3%
}},~$and the resonances are $s_{1}=-1,~s_{2}=\frac{4}{3},~s_{3}=\frac{2}{3}%
~,~s_{4}=\frac{1}{3}.~$From the latter we extract the information that the
step of the series is $\frac{1}{3}$, meaning that the solution is
\begin{equation}
\mu \left( t\right) =\mu _{0}t^{-2}+\sum_{I}^{+\infty }\mu _{I}t^{-2+\frac{I%
}{3}}, \ex\left( t\right) =\nu _{0}t^{-\frac{2}{3}%
}+\sum_{J}^{+\infty }\nu _{J}t^{-\frac{2}{3}+\frac{J}{3}}.  \label{ll.19}
\end{equation}

We substitute the solution \eqref{ll.19} in the system \eqref{2.14}, and we calculate the parameters $%
\mu _{0},\mu _{I}$ and $\nu _{0},\nu _{J}$. We find$~\mu _{I}=\left( \frac{4%
}{3}C_{2},\mu _{2},0,\frac{17C_{1}-16C_{2}}{28C_{1}C_{2}}\left( \mu
_{2}\right) ^{2},0,\ldots\right) $, and,$~\nu _{I}=\left( 0,\frac{C_{1}+4C_{2}}{%
4C_{1}C_{2}},0,0,0, \ldots\right)$, where $\mu _{2},~\nu _{0},~$are the constants of integration. The third constant of integration is in the next coefficient.
Therefore we say that for $\gamma $ a rational number, for which the dominant
behavior and the resonances are rational numbers, the field equations (\ref%
{ll.08}), (\ref{ll.09}) pass the singularity test.

Now,  as we have proved that the field equations are integrable, that is, that
there exists an actual solution, the evolution of the field equations will be
studied in the following by using dynamical system tools.

\section{Evolution on  a phase-plane}

In order to perform the dynamical-system analysis we need to introduce suitable
 normalized variables that will reduce the system  to a dimensionless form
\cite{WE,Coley:2003mj,Copeland:1997et}.

\subsection{Dynamical system at the finite region}
\label{finite_D_norm}

We follow the approach of Section 5.2 of \cite{Coley:2015qqa}, and define:
 \begin{equation}\label{KS_vars_1}
 x=\frac{\sqrt{\mu}}{D}, y=\frac{\sqrt{3}\sigma}{D}, z=\frac{\sqrt{K}}{D},
Q=\frac{\theta}{\sqrt{3}D},
 \end{equation}
 \noindent
 where:\\
 \begin{equation}
 D=\sqrt{K+\frac{\theta^2}{3}},
 \end{equation}
and the new time variable $f':=\frac{d f}{d\tau}\equiv\frac{1}{D} \dot f.$

The variables \eqref{KS_vars_1} are related through the constraints
\begin{subequations}\label{constraints_KS_1}
 \begin{align}
(1-C_2) Q^2-C_1 y^2+x^2&=1, \label{constraints_KS_1a}\\
Q^2+z^2&=1.
 \end{align}
\end{subequations}
From the equations \eqref{constraints_KS_1} it follows that $Q$ and $z$ are bounded in
the
intervals $Q\in [-1, 1], \; z \in [0, 1]$ (for expanding universes $Q\geq 0$). $x$ and $y$
are bounded for $C_1<0$, i.e., for $c_\sigma\leq \frac{1}{2}.$ Otherwise, they can be unbounded. That is $x\rightarrow \infty, y\rightarrow \infty$ while maintaining $-C_1 y^2+x^2$ bounded with the rough estimate $|-C_1 y^2+x^2|\leq 1+|1-C_2|$.

The restrictions \eqref{constraints_KS_1} allow us to eliminate two variables, say $x$ and
$z.$  This
leads to the following 2-dimensional dynamical system \cite{Coley:2015qqa}:
\begin{subequations}
\label{KS_gen_syst}
  \begin{align}
&y'=\frac{Q y \left(C_2 \left((3 \gamma -2)
   Q^2-4\right)-(3 \gamma -2)
   \left(Q^2-1\right)\right)}{2 \sqrt{3}
   C_2}-\frac{Q^2-1}{\sqrt{3}
   C_1}\nonumber \\ & +\frac{\sqrt{3} (\gamma -2) C_1 Q
   y^3}{2 C_2}-\frac{\left(Q^2-1\right)
   y^2}{\sqrt{3}},\\
&Q'=\frac{\left(Q^2-1\right) \left(C_2 Q (3
   \gamma  Q-2 (Q+y))+3 (\gamma -2) C_1
   y^2-(3 \gamma -2)
   \left(Q^2-1\right)\right)}{2 \sqrt{3}
   C_2},
	\end{align}
\end{subequations}
defined in the invariant set:
$\left\{(y,Q): -C_2 Q^2-C_1 y^2+Q^2\leq 1, Q\in [-1, 1]\right\}.$
			
	\begin{table}
		\renewcommand{\arraystretch}{1.5}
		\resizebox{\columnwidth}{!}
		{%
			\begin{tabular}{|l|ccc|}
			\hline
				Label & Coordinates: $(y,Q)$& Existence	& Eigenvalues \\	
\hline
				$P_{1}$ & $(0, -1)$	& $C_2\geq 0$ &
$\frac{1}{2} \sqrt{3} (2-\gamma),\frac{2-
3 \gamma }{\sqrt{3}}$   \\[1mm]
				\hline
				$P_{2}$ & $(0, 1)$	& $C_2\geq 0$ &
$-\frac{1}{2} \sqrt{3} (2-\gamma),-\frac{
2-3 \gamma}{\sqrt{3}}$   \\[1mm]				
				\hline
				$P_3$ & $\left(-\sqrt{-\frac{C_2}{C_1}}, -1\right)$  & $C_1<0,  C_2\geq 0$ or $C_1>0, C_2\leq 0$ & $-\sqrt{3}(2- \gamma),
				-\frac{4}{\sqrt{3}}+\frac{2}{\sqrt{3}}\sqrt{-\frac{C_2}{C_1}}$	\\[1mm]				
				\hline
				$P_4$ & $\left(\sqrt{-\frac{C_2}{C_1}}, -1 \right)$ & $C_1<0,  C_2\geq 0$ or $C_1>0, C_2\leq 0$ & $-\sqrt{3}(2- \gamma), -\frac{4}{\sqrt{3}}
				-\frac{2}{\sqrt{3}}\sqrt{-\frac{C_2}{C_1}}$	\\[1mm]				
				\hline
				$P_5$ & $\left(-\sqrt{-\frac{C_2}{C_1}}, 1\right)$ & $C_1<0,  C_2\geq 0$ $C_1>0, C_2\leq 0$ & $\sqrt{3}(2-
\gamma) , \frac{4}{\sqrt{3}}+\frac{2}{\sqrt{3}}\sqrt{-\frac{C_2}{C_1}}$	\\[1mm]	\hline
				$P_6$ & $\left(\sqrt{-\frac{C_2}{C_1}}, 1\right)$ & $C_1<0,  C_2\geq 0$ or $C_1>0, C_2\leq 0$ & $\sqrt{3}(2-\gamma) ,
				\frac{4}{\sqrt{3}}-\frac{2}{\sqrt{3}}\sqrt{-\frac{C_2}{C_1}}$	\\[1mm]				
				\hline
				$P_7$ & $\left(\frac{C_2}{C_3},-\frac{2|C_1|}{C_3}\right)$
				& $C_1<0, C_2\leq 0$ or & \\
				&& $C_1<0,  C_2 \geq -4 C_1$ or & \\
				&& $C_1>0, -4 C_1\leq C_2 \leq 0$   & $\frac{4 C_1+C_2}{\sqrt{3}C_3},
				\frac{2 \left((3\gamma -2) C_1+C_2\right)}{\sqrt{3}C_3}$	\\[1mm]				
				\hline
				$P_8$ & $\left(-\frac{C_2}{C_3},\frac{2|C_1|}{C_3}\right)$
				& $C_1<0, C_2\leq 0$ or & \\
				&& $C_1<0,  C_2 \geq -4 C_1$ or & \\
				&& $C_1>0, -4 C_1\leq C_2 \leq 0$     & $-\frac{4 C_1+C_2}{\sqrt{3}C_3},
				-\frac{2 \left((3\gamma -2) C_1+C_2\right)}{\sqrt{3}C_3}$	\\[1mm]				
					\hline
				$P_9$ & $\left(\frac{2-3 \gamma }{C_4}, -\frac{2}{C_4}\right)$ & $0\leq \gamma \leq \frac{2}{3}, C_1\leq 0, C_2\geq (2-3\gamma)C_1$ or & \\
							&& $\frac{2}{3}\leq \gamma<2, C_1\geq 0, C_2\geq (2-3\gamma)C_1$ & $-\frac{\sqrt{3} (\gamma -2)+C_5}{2C_4},
							-\frac{\sqrt{3} (\gamma -2)-C_5}{2C_4}$	\\[1mm]				
					\hline
				$P_{10}$ & $\left(-\frac{2-3 \gamma }{C_4}, \frac{2}{C_4}\right)$ & $0\leq \gamma \leq \frac{2}{3}, C_1\leq 0, C_2\geq (2-3\gamma)C_1$ or & \\
							&& $\frac{2}{3}\leq \gamma<2, C_1\geq 0, C_2\geq (2-3\gamma)C_1$   & $\frac{\sqrt{3} (\gamma -2)+C_5}{2C_4}, \frac{\sqrt{3} (\gamma -2)-C_5}{2C_4}$	\\
[1mm]				
					\hline
			\end{tabular}
		}
		\caption{\label{Tab99} Critical points of the system \eqref{KS_gen_syst}. We use the
notations $C_3=\sqrt{C_1\left(-C_2^2-4 C_1\left(C_2-1\right)\right)}$ and
$C_4=\sqrt{4-3 (\gamma -2) (3 \gamma -2) C_1},$ and
$C_5=\frac{\sqrt{(\gamma -2) \left(8 (2-3\gamma )^2 C_1+(27 \gamma -22)C_2\right)}}{\sqrt{C_2}}.$ We have assumed that $0\leq \gamma < 2$.}
	\end{table}

The  usual volume deceleration parameter, $q=-1-3 \dot{\theta}/\theta^2$, is
given by
				\begin{equation}
				q Q^2=\frac{(3 \gamma -2) \left(C_2 Q^2-Q^2+1\right)+3 (\gamma -2) C_1 y^2}{2 C_2}.
				\end{equation}

This system was deduced in \cite{Coley:2015qqa}, but the analysis was done for a special choice of parameters.
 Now, we will discuss the system in detail, without specifying the values of the \ae ther parameters $c_a, c_\theta$ and $c_\sigma$. As applications, we select values for the \ae ther parameters which are consistent with current constraints, generating a very rich phenomenology as we will discuss shortly. Particularly, the results found in \cite{Coley:2015qqa} are re-obtained as particular cases. But first, let us comment on the
stability
conditions of the critical points of the system \eqref{KS_gen_syst} given in Table
\ref{Tab99}. Furthermore, at Table \ref{grpoints} the coordinates of the critical points
at the limit of GR,~$C_{2}=-C_{1}=1$,  are presented. 
 
\begin{table}[tbp] \centering%
\begin{tabular}{|c|c|c|}
\hline
Point & Gen. Relat. $(y,Q)$ & Asymptotic behavior\\ \hline
$P_{1}$ & $\left( 0,-1\right)$ & $\mu= \frac{\widetilde{\mu_0}}{\Delta t^2},K=
   \widetilde{K_0} \Delta t^{-\frac{4}{3\gamma}},\ex=\widetilde{\ex_0} \Delta t^{-\frac{2}{3\gamma}},\ell=\widetilde{\ell_0}\Delta t^{\frac{2}{3\gamma}}$. \\ 
	 & & A point--like singularity as $\ell\rightarrow 0$.  \\ \hline
$P_{2}$ & $\left( 0,1\right) $ & $\mu= \frac{\widetilde{\mu_0}}{\Delta t^2},K=
   \widetilde{K_0} \Delta t^{-\frac{4}{3\gamma}},\ex=\widetilde{\ex_0} \Delta t^{-\frac{2}{3\gamma}},\ell=\widetilde{\ell_0}\Delta t^{\frac{2}{3\gamma}}$. \\ 
	 & & A point--like singularity as $\ell\rightarrow 0$.  \\ \hline
$P_{3}$ & $\left( -1,-1\right) $ &  $\mu= \frac{\widetilde{\mu_0}}{\Delta t^\gamma},K=
   \widetilde{K_0} \Delta t^{-\frac{4}{3}},\ex=\widetilde{\ex_0} \Delta t^{\frac{1}{3}},\ell=\widetilde{\ell_0}\Delta t^{\frac{1}{3}}$. \\
	& & Cigar singularity as $\ell\rightarrow 0$.\\ \hline
$P_{4}$ & $\left( 1,-1\right) $ & $\mu= \frac{\widetilde{\mu_0}}{\Delta t^\gamma},K=
   \widetilde{K_0},\ex=\widetilde{\ex_0} \Delta t^{-1},\ell=\widetilde{\ell_0}\Delta t^{\frac{1}{3}}$. \\
	& &  A pancake singularity as $\ell\rightarrow 0$. \\ \hline
$P_{5}$ & $\left( -1,1\right) $ & $\mu= \frac{\widetilde{\mu_0}}{\Delta t^\gamma},K=
   \widetilde{K_0},\ex=\widetilde{\ex_0} \Delta t^{-1},\ell=\widetilde{\ell_0}\Delta t^{\frac{1}{3}}$.   \\
	& &  A pancake singularity as $\ell\rightarrow 0$. \\ \hline
$P_{6}$ & $\left( 1,1\right) $ & $\mu= \frac{\widetilde{\mu_0}}{\Delta t^\gamma},K=
   \widetilde{K_0} \Delta t^{-\frac{4}{3}},\ex=\widetilde{\ex_0} \Delta t^{\frac{1}{3}},\ell=\widetilde{\ell_0}\Delta t^{\frac{1}{3}}$. \\
	& & Cigar singularity as $\ell\rightarrow 0$.\\ \hline
$P_{9}$ & $\left( \frac{2-3\gamma }{%
3\gamma -4},-\frac{2}{3\gamma -4}\right)$,\; $\gamma < \frac{2}{3}$ & $\mu= \frac{\widetilde{\mu_0}}{\Delta t^2},K=
  \frac{\widetilde{K_0}}{\Delta t^2},\ex=\widetilde{\ex_0} \Delta t^{2-\frac{2}{\gamma}},\ell=\widetilde{\ell_0}\Delta t^{\frac{2}{3\gamma}}$. \\
	& &  A point--like singularity as $\ell\rightarrow 0$.\\ \hline
$P_{10}$ & $\left( -\frac{2-3\gamma }{%
3\gamma -4},-\frac{2}{3\gamma -4}\right)$,\; $\gamma < \frac{2}{3}$ & $\mu= \frac{\widetilde{\mu_0}}{\Delta t^2},K=
  \frac{\widetilde{K_0}}{\Delta t^2},\ex=\widetilde{\ex_0} \Delta t^{2-\frac{2}{\gamma}},\ell=\widetilde{\ell_0}\Delta t^{\frac{2}{3\gamma}}$. \\
	& &  A point--like singularity as $\ell\rightarrow 0$.\\ \hline
\end{tabular}%
\caption{\label{grpoints} Critical points of the system \eqref{KS_gen_syst} at the limit of GR,
that is, $C_{2}=-C_{1}=1$.}
\end{table}%

		The critical points $P_1$ and $P_2$ exist for $C_2\geq 0$. $P_1$ is a source and $P_2$ is a sink for $C_2\geq 0$, and $0\leq \gamma <\frac{2}{3}$. They are nonhyperbolic for $\gamma=\frac{2}{3}$ and saddles otherwise. The deceleration factor evaluated at the critical points $P_1$ and $P_2$ is given by $q=\frac{3 \gamma }{2}-1$. On the other hand, the sign of $Q$ means expansion if it is positive, and contraction if it is negative. Thus, $P_1$ (respectively, $P_2$) corresponds to solutions with decelerated contraction (respectively, decelerated expansion) for $\gamma>\frac{2}{3}$ and accelerated contraction (respectively, decelerated contraction) for $0\leq \gamma <\frac{2}{3}$.

The critical points $P_3, P_4, P_5$ and $P_6$ exist for $C_1<0, C_2\geq 0$ or $C_1>0, C_2\leq 0$.
 $P_3$ is:
	  \begin{enumerate}
		\item nonhyperbolic for $C_2\neq 0, C_2=-4 C_1$,
	  \item a sink for
		   \begin{enumerate}
			  \item $0\leq \gamma <2, C_1<0, 0\leq C_2<-4 C_1$, or
				\item $0\leq \gamma <2, C_1>0, -4C_1<C_2\leq 0$,
				\end{enumerate}
				or
	  \item a saddle otherwise.
		\end{enumerate}
	$P_4$ is a sink whenever exists. That is, for $C_1<0,  C_2\geq 0$ or $C_1>0, C_2\leq 0$.\\
	$P_5$ is a source whenever exists. That is, for $C_1<0,  C_2\geq 0$ or $C_1>0, C_2\leq 0$.\\
	$P_6$ is:
	  \begin{enumerate}
		 \item nonhyperbolic for $C_2\neq 0, C_2=-4 C_1$,
		 \item a source for $0\leq \gamma <2, C_1<0, 0\leq C_2<-4 C_1$, or $0\leq \gamma <2, C_1>0, -4C_1<C_2\leq 0$,
	   \item a saddle otherwise.
		\end{enumerate}
		The deceleration parameter evaluated at the critical points $P_3$ to $P_6$ is
given by $q=2$. Thus, $P_3$ and $P_4$ represent decelerated expanding stiff-like fluid solutions, while $P_5$ and $P_4$  corresponds to decelerated stiff-like contracting solutions. \\
The critical points $P_7$ and $P_8$ exist for
	\begin{enumerate}
	 \item $C_1<0, C_2\leq 0$ or
	 \item $C_1<0,  C_2 \geq -4 C_1$ or
	 \item $C_1>0, -4 C_1\leq C_2 \leq 0$.
	\end{enumerate}
	Since GR is recovered for the specific choice of parameters $C_1=-1, C_2=1$
it follows that these points are not allowed in GR. To our knowledge, the existence of these non-GR anisotropic states $P_7$ and $P_8$ was first partially proved in \cite{Coley:2015qqa} for the specific case $c_\sigma=\frac{1}{2}(1- c^2)  \geq 0,
 		c_a =  -\frac{d}{(1+d)} c_\sigma \leq 0, c_\theta=0$, where $c$ and $d$ are constants. That is for $C_1=-c^2\leq 0, C_2=1, c_a =  -\frac{d(1- c^2)}{2(1+d)} $. They are a non-trivial consequence of the presence of a non-zero
Lorentz-violating vector field. Now, let us discuss on their stability. \\
	$P_7$ (resp. $P_8$) is:
	\begin{enumerate}
	\item nonhyperbolic for:
	\begin{enumerate}
	\item $C_1>0, C_2=-4 C_1, 0\leq \gamma <2$ or
	\item $C_1>0, -4 C_1<C_2\leq 0, \gamma
   =\frac{2 C_1-C_2}{3 C_1}$.
	\end{enumerate}
		\item a sink (resp. a source) for
		  \begin{enumerate}
			\item $0\leq \gamma\leq\frac{2}{3}, C_1<0, C_2<(2-3\gamma)C_1$, or
			\item $\frac{2}{3}<\gamma<2, C_1<0, C_2\leq 0$.
			\end{enumerate}
	\item a source (resp. a sink) for
	    \begin{enumerate}
			\item $0\leq \gamma< 2, C_1<0, C_2>-4 C_1$, or
			\item $\frac{2}{3}<\gamma <2, C_1>0, (2-3\gamma ) C_1<C_2\leq 0$.
			\end{enumerate}
	\item a saddle otherwise.
	\end{enumerate}
	
	The deceleration parameter evaluated at the critical points is given by $q=-\frac{C_2}{2 C_1}$, and thus, the critical point represents:
					\begin{enumerate}
					\item an accelerated solution for
					      \begin{enumerate}
								\item $0\leq \gamma <2, C_1<0, C_2<0$ [$P_7$ is a late-time accelerated solution,  dark energy, sink; $P_8$ is an early-time accelerated,
inflationary solution, source].
								\end{enumerate}
								
          \item a decelerated solution for
                 \begin{enumerate}
								  \item $0\leq \gamma <2, -4 C_1\leq C_2<0$ [$P_7$ and $P_8$ are saddles] or
								  \item $0\leq \gamma <2, C_1<0, C_2\geq -4C_1$ [$P_7$ is a sink; $P_8$ is a source].
								 \end{enumerate}
					\end{enumerate}

	$P_9$ and $P_{10}$ exist for
	\begin{enumerate}
	 \item $0\leq \gamma \leq \frac{2}{3}, C_1\leq 0, C_2\geq (2-3\gamma)C_1$ or
	 \item $\frac{2}{3}\leq \gamma<2, C_1\geq 0, C_2\geq (2-3\gamma)C_1$
	\end{enumerate}
	
	$P_9$ (resp. $P_{10}$) is nonhyperbolic for
	 \begin{enumerate}
	 \item $0\leq \gamma <\frac{2}{3},  C_1<0, C_2=(2-3 \gamma )C_1$, or
	 \item $\gamma=\frac{2}{3}, C_2>0$, or
   \item $\frac{2}{3}<\gamma <2, C_1>0, C_2=(2-3 \gamma ) C_1$.
	 \end{enumerate}
	
The trace and the determinant of the Jacobian matrix evaluated at $P_{9,10}$ are $\tau=\pm\frac{2-\gamma}{\sqrt{((8-3
   \gamma ) \gamma -4)   C_1+\frac{4}{3}}}, \delta=\frac{2 (\gamma -2)
   (3 \gamma -2) \left((3 \gamma -2)
   C_1+C_2\right)}{C_2 \left(3 (\gamma -2)
   (3 \gamma -2) C_1-4\right)}$, respectively. Thus, $P_9$ (resp. $P_{10}$) is:
	  \begin{enumerate}
		 \item a source (resp. a sink) for
		  \begin{enumerate}
			 \item $0\leq \gamma <\frac{2}{3}, C_1<0, (2-3 \gamma ) C_1<C_2<0$, or
		   \item $\frac{2}{3}<\gamma <2, C_1\geq 0, C_2>0$.
			\end{enumerate}
	\item a saddle for
	  \begin{enumerate}
		\item $0\leq \gamma <\frac{2}{3}, C_1\leq 0, C_2>0$, or
		\item $\frac{2}{3}<\gamma <2, C_1>0, (2-3 \gamma)C_1<C_2<0$.
		\end{enumerate}
	 \end{enumerate}

The deceleration factor evaluated at the critical point is given by $q=\frac{3 \gamma
}{2}-1$. Thus, the solutions are decelerated for $\gamma>\frac{2}{3}$ and accelerated for $0\leq \gamma<\frac{2}{3}$.

\subsection{Dynamical system at infinity.}

Due to the fact that the dynamical system \eqref{KS_gen_syst} is
non-compact (along the $y$-
direction, since $Q\in[-1,1]$ is bounded), there could be features in the asymptotic
regime which
are non trivial for the global dynamics. Thus, in order to complete the analysis of the
phase space
we will now extend our study using in place of $y$ a variable that remains finite in the asymptotic regime.
The new variable $v=y/\sqrt{1+y^2}$ ensures that the regimes
$y\rightarrow \pm \infty$ are mapped onto $v\rightarrow \pm 1$.
	
		The field equations becomes	
{\small
			 \begin{subequations}
		\label{infinity}
		\begin{align}
	&\frac{d v}{d T}=\frac{(3 \gamma -2) \left(C_2-1\right) Q^3 v \left(v^2-1\right)^2}{2 \sqrt{3}
   C_2}-\frac{\left(Q^2-1\right) \left(1-v^2\right)^{3/2} \left(C_1 v^2-v^2+1\right)}{\sqrt{3} C_1} \nonumber \\
	& -\frac{Q v \left(v^2-1\right) \left(3 (\gamma -2) C_1 v^2-\left(v^2-1\right) \left(3 \gamma -4 C_2-2\right)\right)}{2\sqrt{3} C_2},\\
	&\frac{d Q}{d T}=-\frac{(3 \gamma -2) \left(C_2-1\right) Q^4 \left(v^2-1\right)}{2 \sqrt{3} C_2}+\frac{Q^2
   \left(3 (\gamma -2) C_1 v^2+(3 \gamma -2) \left(C_2-2\right) \left(v^2-1\right)\right)}{2 \sqrt{3}
   C_2} \nonumber \\ & +\frac{(3 \gamma -2) \left(v^2-1\right)-3 (\gamma -2) C_1 v^2}{2 \sqrt{3} C_2}-\frac{1}{3}
   \left(Q^2-1\right) Q v \sqrt{3-3 v^2},
		\end{align}
		\end{subequations}
		}
		where we have used the time rescaling  $\frac{d f}{d T}= (1-v^2) \frac{d
f}{d \tau}$. Hence, the phase space transforms to
\begin{align}
\left\{(v,Q): (1-C_2) Q^2-\frac{C_1 v^2}{1-v^2}\leq 1, v\in [-1, 1], Q\in
[-1, 1]\right\}.
 \end{align}
		
		Apart from the critical points analogous to $P_1$-$P_{10}$ in the
$(v,Q)$ plane we have the addition
of four critical points with coordinates $Q_{1,2}: (v,Q)=(\mp 1 , -1)$ and $Q_{3,4}:
(v,Q)=(\mp 1, 1)$.
 The eigenvalues of the linearization of \eqref{infinity} around each of these fixed
points are $\left\{-\frac{\sqrt{3} (\gamma -2) C_1}{C_2},\frac{\sqrt{3} (\gamma -2) C_1}{C_2}\right\}$. Thus, they are always saddles for
$\gamma\neq 2$ and $C_1\neq 0$.
	
	At the invariant set $v=\pm 1$ we have the solution
	\begin{equation}
	Q(T)= -\tanh \left(\alpha-\frac{\sqrt{3} (2-\gamma) C_1 T}{2 C_2}\right),
	\end{equation} where $\alpha$ is an integration constant such that $\alpha=0$ for the solution with initial condition $Q(0)=0$.
	This implies, for example, that the solutions with zero expansion but very high shear $\sigma/\sqrt{3 K+\theta^2}\rightarrow \pm \infty$, i.e., with $v\rightarrow\pm 1$, satisfy $Q\rightarrow \pm 1$, depending of the sign of $C_1/C_2$. That is, they connect accelerated expansion era with a decelerated expansion era, or viceversa. In GR, where  $C_1/C_2=-1$, and for $0\leq \gamma<2$, it follows that $Q\rightarrow +1$ as $T\rightarrow -\infty$ and  $Q\rightarrow -1$ as $T\rightarrow +\infty$, and we get early-time expanding solutions and late-time contracting solution. When we depart from the invariant sets $v=\pm 1$, these
solutions, with extremely high anisotropy, are of saddle type.
	As long as $y$ is infinite, $|-3 c_\theta Q^2+x^2|$ is infinite too, because the
restriction \eqref{constraints_KS_1a} has to be satisfied. Additionally, since $Q\rightarrow
\pm 1$ according of the sign of $C_1/C_2$, it follows that extremely high anisotropic solutions also have zero curvature ($K\rightarrow
0$), and large matter energy density ($|x|\rightarrow \infty$) in comparison with the Hubble
scalar.
		
\section{Exact solutions at the critical points}

At the critical points the equations \eqref{ll.01}, \eqref{ll.05}, \eqref{ll.04}, and the new equation $\dot\ell/\ell=\theta/3$ (that defines the length scale $\ell$
 along the flow lines) become:
\begin{subequations}
\begin{align}
& \frac{d K}{d\tau}=-\frac{2 (Q^*+y^*) K}{\sqrt{3}}\implies K=K_0 e^{-\frac{2 (Q^*+y^*)\tau}{\sqrt{3}}} ,\\
& \frac{d\ex}{d\tau}=-\frac{\ex (Q^*-2 y^*)}{\sqrt{3}}\implies \ex={\ex}_0 e^{-\frac{(Q^*-2 y^*)\tau}{\sqrt{3}}},\\
& \frac{d\ell}{d\tau}=\frac{Q^*  \ell}{\sqrt{3}}\implies \ell=\ell_0 e^{\frac{Q^*\tau}{\sqrt{3}}}\\
& \frac{d \mu}{d\tau}=-\sqrt{3} \gamma  Q^* \mu \implies \mu=\mu_0 e^{-\sqrt{3} \gamma  Q^* \tau},
\end{align}
\end{subequations}
 where the star-upperscript denotes the evaluation at a specific critical point.
In order to express the above determined functions of $\tau$ in
terms of the comoving time variable $t$, we solve the system:
\begin{equation}
\label{inverse}
\frac{d\tau}{ dt}=D,\quad
\frac{d D}{ dt}=D^2 \Upsilon^*,
\end{equation}
where
$\Upsilon^*=\frac{\left(C_2 \left(2 \left({Q^*}^2-1\right) (Q^*+y^*)-3 \gamma {Q^*}^3\right)-3 (\gamma -2) C_1 Q^* {y^*}^2+(3\gamma -2) Q^*
   \left({Q^*}^2-1\right)\right)}{2 \sqrt{3}C_2}$.\\
	Solving equations \eqref{inverse} (with initial conditions $D(t_0) = D_0$ and $\tau(t_0) = 0$) we obtain
		\begin{equation}
	\tau = \frac{\ln \left(\frac{1}{1- D_0 \Upsilon^* (t-t_0)}\right)}{\Upsilon^*},\quad
	D=\frac{D_0}{1-D_0 \Upsilon^*(t-t_0)}.
	\end{equation}
Hence, we have
\begin{subequations}
\begin{align}
&  \ex= {\ex}_0 (1-D_0 \Upsilon^* (t-t_0))^{\frac{Q^*-2 y^*}{\sqrt{3} \Upsilon^*}}, \\
&	K= K_0 (1-D_0 \Upsilon^* (t-t_0))^{\frac{2 (Q^*+y^*)}{\sqrt{3}\Upsilon^*}},\\
&	\ell= \ell_0 (1-D_0 \Upsilon^* (t-t_0))^{-\frac{Q^*}{\sqrt{3} \Upsilon^*}},\\
&\mu = \mu_{0} (1-D_0 \Upsilon^*   (t-t_0))^{\frac{\sqrt{3} \gamma  Q^*}{\Upsilon^*}}.
\end{align}
\end{subequations}
For illustration we present the asymptotics for the new points $P_7$ and $P_8$. Redefining some constants, we obtain for $P_7$ the solution
\begin{subequations}
\label{approx:sol}
\begin{align}
&\ex=\widetilde{\ex_{0}} (\Delta t)^{2-3p}, \\
&K=\widetilde{K_0} (\Delta t)^{-2},\\
&\ell=\widetilde{\ell_0} \left(\Delta t\right)^{p},\\
& \mu=\widetilde{\mu_{0}}\left(\Delta t\right)^{-3\gamma p},
\end{align}
\end{subequations}
where $p=-\frac{2  C_1}{C_2-2 C_1}$. While for $P_8$ we have similar behavior.
 From the power-law solution of the volume at both points, we observe that when the power $p>0$ the universe admits a Big Bang or a Big Crunch. On the other hand for
$p<0$ at the limit $\Delta t\rightarrow 0$, the volume becomes infinite. Taking into the account the existence conditions for $P_{7,8}$ we find that for $C_1<0, C_2<2 C_1$ the universe does not have a Big Bang or a Big Crunch solution. In this case the point $P_7$ is always a sink and $P_8$ is always a source.\\
The expressions \eqref{approx:sol} are approximate solutions of \eqref{system_1} up to order $\mathcal{O}(\Delta t^{-3 \gamma p})=\mathcal{O}(\ell^{-3\gamma})$. The errors terms tend to zero when $\Delta t\rightarrow +\infty$ for $p>0$ and when $\Delta t\rightarrow 0$ for $p<0$. Note that the error terms are proportional to the matter energy density. Assume that the matter density is constant and defines an order parameter, say $\mu=\mathcal{O}(\epsilon)^2$, with $\epsilon\ll 1$. Then, we find the approximate solution
\begin{subequations}
\begin{align}
&\ex=\widetilde{\ex_{0}} (\Delta t)^{2-3 p}+\frac{\widetilde{\ex_{0}} (3 p-2) \epsilon  (\Delta t)^{1-3 p}}{3 p}+\mathcal{O}(\epsilon)^2,\\
&K=\widetilde{K_0}(\Delta t)^{-2}+\widetilde{K_0} \epsilon  \left(\frac{\alpha  (\Delta t)^{-3 p-1}}{3 p-1}+\frac{2}{3 p}(\Delta t)^{-3}\right)+\mathcal{O}(\epsilon)^2,\\
&\ell=\widetilde{\ell_0} (\Delta t)^p+\widetilde{\ell_0} \epsilon  \left(\frac{\alpha  (\Delta t)^{1-2 p}}{3-9 p}-\frac{1}{3}
  (\Delta t)^{p-1}\right)+\mathcal{O}(\epsilon)^2,\label{5.6c}
\end{align}
\end{subequations}
where $\alpha$ is an integration constant.
Thus,
\begin{subequations}
\label{expressions_5.7}
\begin{align}
&\theta=3 p(\Delta t)^{-1}+\epsilon  \left(\alpha  (\Delta t)^{-3 p}+(\Delta t)^{-2}\right)+\mathcal{O}(\epsilon)^2,\\
&\sigma=(1-p)(\Delta t)^{-1} +\frac{1}{6} \epsilon  \left(\alpha  (\Delta t)^{-3 p}+\frac{2 (1-p)}{p}(\Delta t)^{-2}\right)+\mathcal{O}(\epsilon)^2
\end{align}
\end{subequations}
Inverting \eqref{5.6c} we get
\begin{equation}
\Delta t= \left(\frac{\ell}{\widetilde{\ell_0}}\right)^{\frac{1}{p}}+\epsilon  \left(\frac{\alpha  \left(\frac{\ell}{\widetilde{\ell_0}}\right)^{3-\frac{2}{p}}}{3 p (3 p-1)}+\frac{1}{3 p}\right)+\mathcal{O}(\epsilon)^2.
\end{equation}
Thus,
\begin{subequations}
\begin{align}
&\frac{\ex}{\widetilde{\ex_{0}}}= \left(\frac{\ell}{\widetilde{\ell_0}}\right)^{\frac{2}{p}-3}+\frac{\alpha  (2-3 p) \epsilon \left(\frac{\ell}{\widetilde{\ell_0}}\right)^{-1/p}}{3 p (3 p-1)}+\mathcal{O}(\epsilon)^2,\\
&\frac{\ey}{\widetilde{\ey_{0}}}=\left(\frac{\ell}{\widetilde{\ell_0}}\right)^{-1/p}+\frac{\epsilon \alpha \left(2   \left(\frac{\ell}{\widetilde{\ell_0}}\right)^{3-\frac{4}{p}}-3 p\left(\frac{\ell}{\widetilde{\ell_0}}\right)^{-3}\right)}{6 p(1-3 p)}+\mathcal{O}(\epsilon)^2.
\end{align}
\end{subequations}
To finish this section we comment briefly about the nature of singularities.\\
To classify the singularities we construct the anisotropy tensor
$\Theta_{a b}=\sigma_{a b}+\frac{1}{3}\theta h_{a b}$, where in the Kantowski-Sachs metric $\Theta_{a b}=\text{diag}(0,\Theta_1,\Theta_2,\Theta_3)$, $\Theta_1=\frac{1}{3}\theta-2\sigma, \Theta_2=\Theta_3=\frac{1}{3}\theta+\sigma$. Each $\Theta_\alpha$ defines a length scale through $\Theta_\alpha=\frac{\dot\ell_\alpha}{\ell_\alpha}$. The different singularity types are distinguished by the behavior of the length scales $\ell_\alpha$ as $\ell\rightarrow 0$ \cite{WE}, which in the Kantowski-Sachs metric satisfies $\ell_1\propto (\ex)^{-1}$, $\ell_2$ and $\ell_3 \propto (\ey)^{-1}$. The singularity is (see section 1.3.4 \cite{WE}):
\begin{itemize}
\item A point: if all the length scales shrink to zero, $\ell_1,\ell_2,\ell_3\rightarrow 0$, as $\ell\rightarrow 0$. For Kantowski-Sachs the condition is $\ex \rightarrow \infty,\ey\rightarrow \infty$ as $\ell\rightarrow 0$.
\item cigar: if two of the length scales shrink to zero, and the third one increases without bound as $\ell\rightarrow 0$. For Kantowski-Sachs the condition is $\ex \rightarrow 0,\ey\rightarrow \infty$ as $\ell\rightarrow 0$.
\item a barrel: if two of the length scales shrink to zero, and the third one tends to a finite value as $\ell\rightarrow 0$. For Kantowski-Sachs the condition is $\ex \rightarrow C \neq 0,\ey\rightarrow \infty$ as $\ell\rightarrow 0$.
\item a pancake: if one of the length scales tend to zero, and two approaches finite values as $\ell\rightarrow 0$.  For Kantowski-Sachs the condition is $\ex \rightarrow \infty, \ey\rightarrow C \neq 0$ as $\ell\rightarrow 0$.
\end{itemize}
Hence, for $P_7$, and assuming $p>0$, this guarantees that the Big-bang/Big-cruch singularities exists, and we can have anisotropic structures such as a barrel, cigar or a pancake, or an isotropic point like structure, depending on the initial conditions on anisotropic shear and matter. For example, assuming initial conditions such that $\alpha=0$, we have that the singularity is point-like for $p>\frac{2}{3}$ and a cigar for  $0<p<\frac{2}{3}$. Now, for $\alpha\neq 0$, and $0<p<\frac{2}{3}$ or $p>\frac{4}{3}$, the singularity is point-like. For $\frac{2}{3}<p<\frac{4}{3}$, $\ex$ exhibits a typical indeterminacy $\infty \cdot 0$,  and $\ey\rightarrow \infty$ as $\ell\rightarrow 0$. Thus, the singularity can be a point, cigar or a barrel. However, the complete analysis of singularities is out the reach of the present research.
\section{Applications}

For the applications we use the parameter choices discussed in the papers
\cite{Jacobson,Barausse:2011pu,Kanno:2006ty} and
summarized in  Appendix A of \cite{Coley:2015qqa}.

\subsection{Case A}\label{CaseA}
\begin{figure}[htb] \centering
\subfigure[$c_\theta=0, c=0.3$ and $\gamma=1$.]{\includegraphics[width=0.45\textwidth]{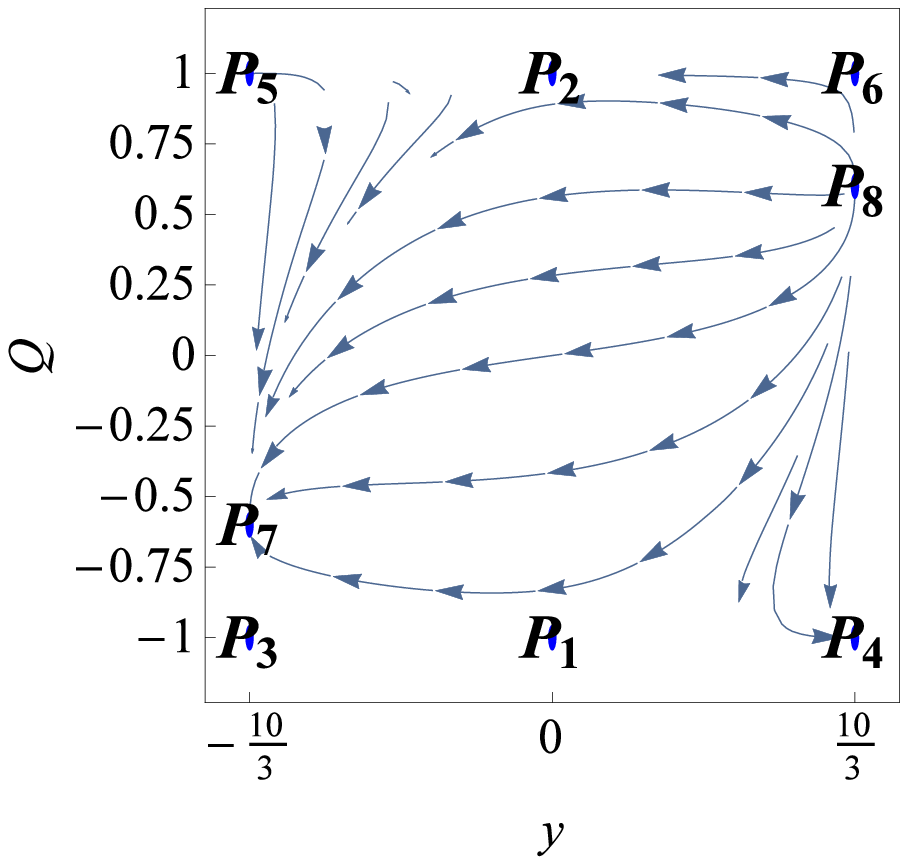}}
\subfigure[$c_\theta=0, c=\sqrt{\frac{3}{5}}$ and $\gamma=1$]{\includegraphics[width=0.45\textwidth]{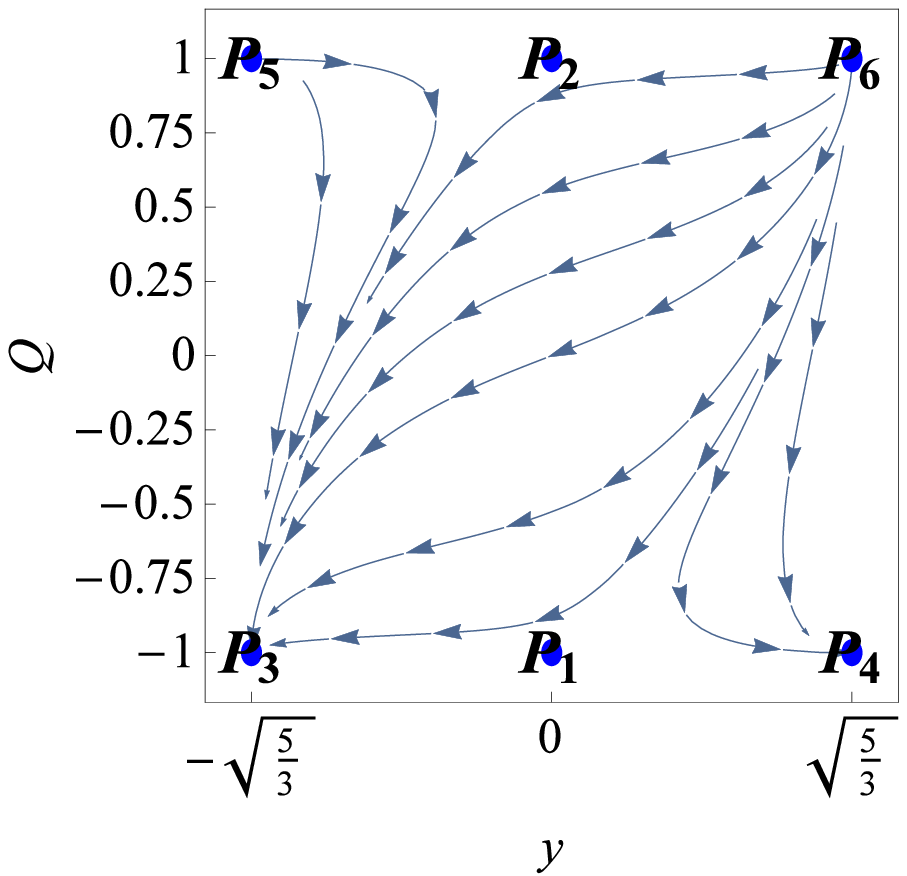}}
\subfigure[$c_\theta=0, c=0.3$ and $\gamma=0$. ]{\includegraphics[width=0.45\textwidth]{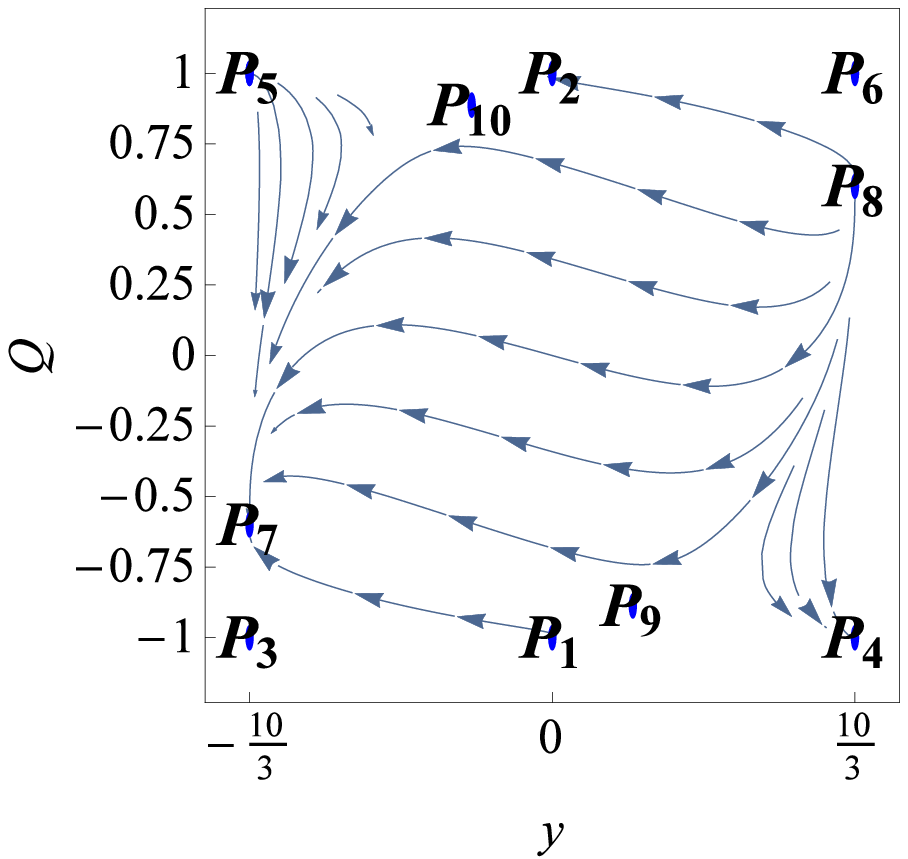}}
\subfigure[$c_\theta=0, c=0.6$ and $\gamma=0$. ]{\includegraphics[width=0.45\textwidth]{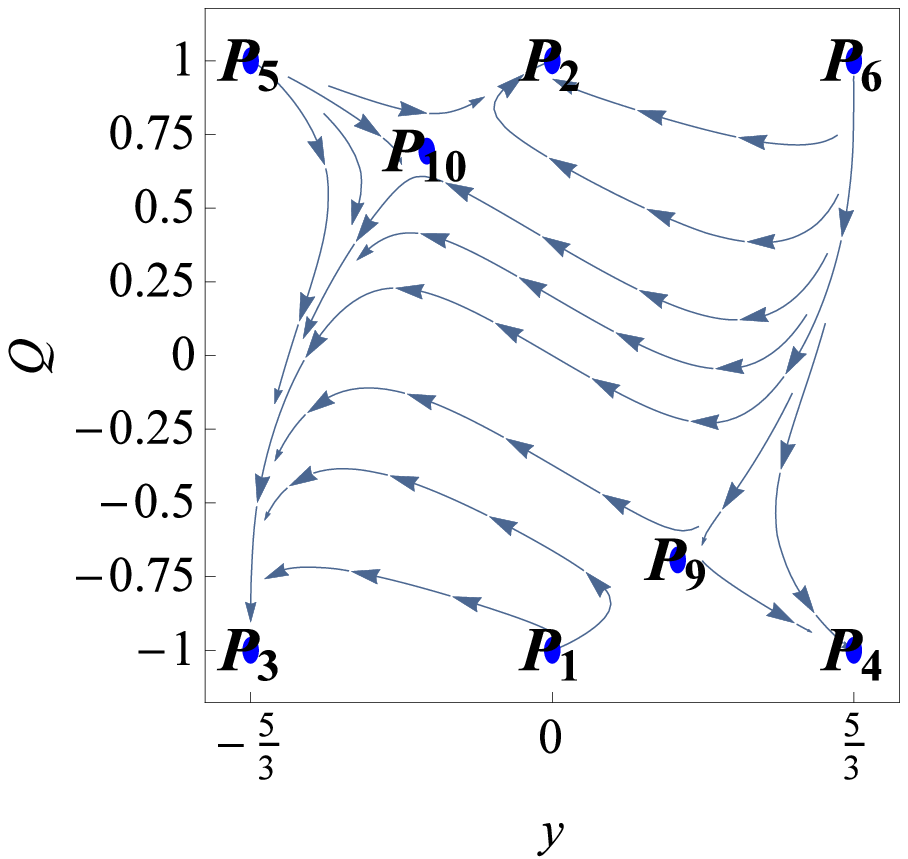}}
\caption{Phase plots of the system \eqref{example1}.} \label{fig:1} \end{figure}

		Let us choose  $c_\sigma=\frac{1}{2}(1- c^2)  \geq 0,
 		c_a =  -\frac{d}{(1+d)} c_\sigma \leq 0, c_\theta=0$. Without the loss of generality we can choose $c>0$.
		In this special case the system \eqref{KS_gen_syst} becomes
		\begin{subequations}
		\label{example1}
		\begin{align}
		& y'=-\frac{\left(c^2 y^2-1\right) \left(3 (\gamma -2) c^2 Q
   y+2 Q^2-2\right)}{2 \sqrt{3} c^2},\\
	  & Q'=-\frac{\left(Q^2-1\right)
   \left(3 c^2 (\gamma -2) y^2-3 \gamma +2 Q y+2\right)}{2
   \sqrt{3}},
		\end{align}
		\end{subequations}
 and the phase space becomes compact $\{(y,Q): -\frac{1}{|c|}\leq y\leq \frac{1}{|c|}, -1\leq Q\leq 1\}.$\\
 This case is not new and it was fully discussed in \cite{Coley:2015qqa}, but using different auxiliary variables. From the mathematical point of view, the simulations displayed in figure  \ref{fig:1} (a), (b), (c), (d), and the simulations displayed in the figures 6, 7, 4, 5
of \cite{Coley:2015qqa}, represent topologically equivalent flows. Indeed, under the rescaling $y\rightarrow y/c, t\rightarrow t/(2\sqrt{3} c), c>0$, we recover the
system (5.27) investigated in \cite{Coley:2015qqa}. The following results are recovered: for $c_\sigma<\frac{1}{2}$ (i.e., $c>0$), when $\gamma < \frac{2}{3}$,
$P_2$ is the unique shear-free, zero curvature (FLRW) inflationary future attractor, and for
$\frac{3}{8}<c_\sigma<\frac{1}{2}$ (i.e., $0<c<\frac{1}{2}$) and $0\leq\gamma<2$ the sources and sinks are, respectively, $P_5$ \& $P_8$ and
$P_4$ \& $P_7$ (as confirmed in figure \ref{fig:1} (a)). All of these sources and sinks are anisotropic and all, except $P_7$, have zero curvature;
the sink  $P_7$ does not have zero curvature. For $c_\sigma<\frac{3}{8}$ (i.e., $c>\frac{1}{2}$)
the points $P_7$ \& $P_8$ do not exist, and the sources and sinks with non-zero shear are $P_5$
and $P_4$, respectively (as confirmed in figure \ref{fig:1} (b)). \footnote{In this paper we changed the labels $P_4$ and $P_5$ in comparison with reference \cite{Coley:2015qqa}.}

Finally, we discuss the special case $\gamma=0$, which corresponds to a Cosmological Constant, due to its cosmological interest. In this case emerge additional source $P_1$ and the additional sink $P_2$, as shown in Figures \ref{fig:1} (c,d), and they correspond to the de Sitter solutions ($q=-1$). The anisotropic solutions $P_9$ and $P_{10}$ become saddles.

\subsection{Case B}
\begin{figure}[htb] \centering
\subfigure[$c_1=-\frac{1}{3}, c_3=\frac{3}{5}, \gamma=1$]{\includegraphics[width=0.45\textwidth]{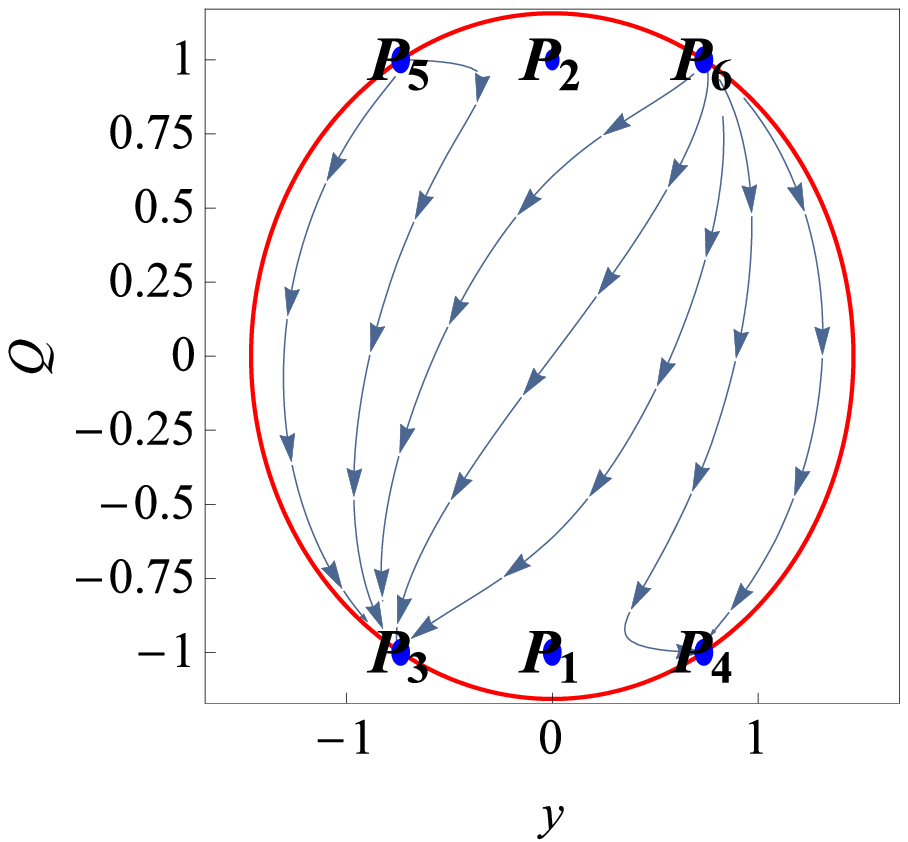}}
\subfigure[$c_1=\frac{1}{4}, c_3=0, \gamma=1$]{\includegraphics[width=0.45\textwidth]{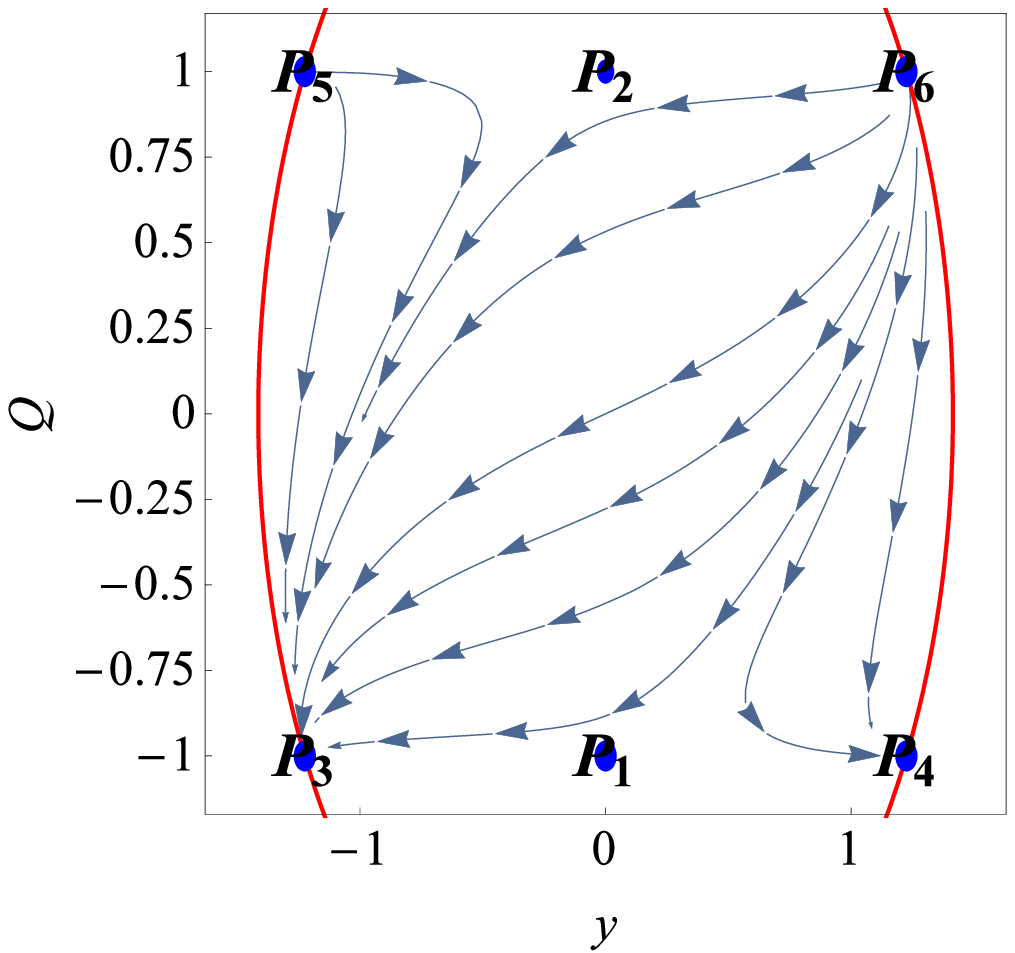}}
\caption{Streamline plot of the vector field  \eqref{example2}. Red (continuous) lines represent the boundary of the phase space.} \label{fig:B} \end{figure}

\begin{figure}[htb] \centering
\subfigure[$c_1=-\frac{1}{2}, c_3=\frac{3}{2}, \gamma=1$]{\includegraphics[width=0.4\textwidth]{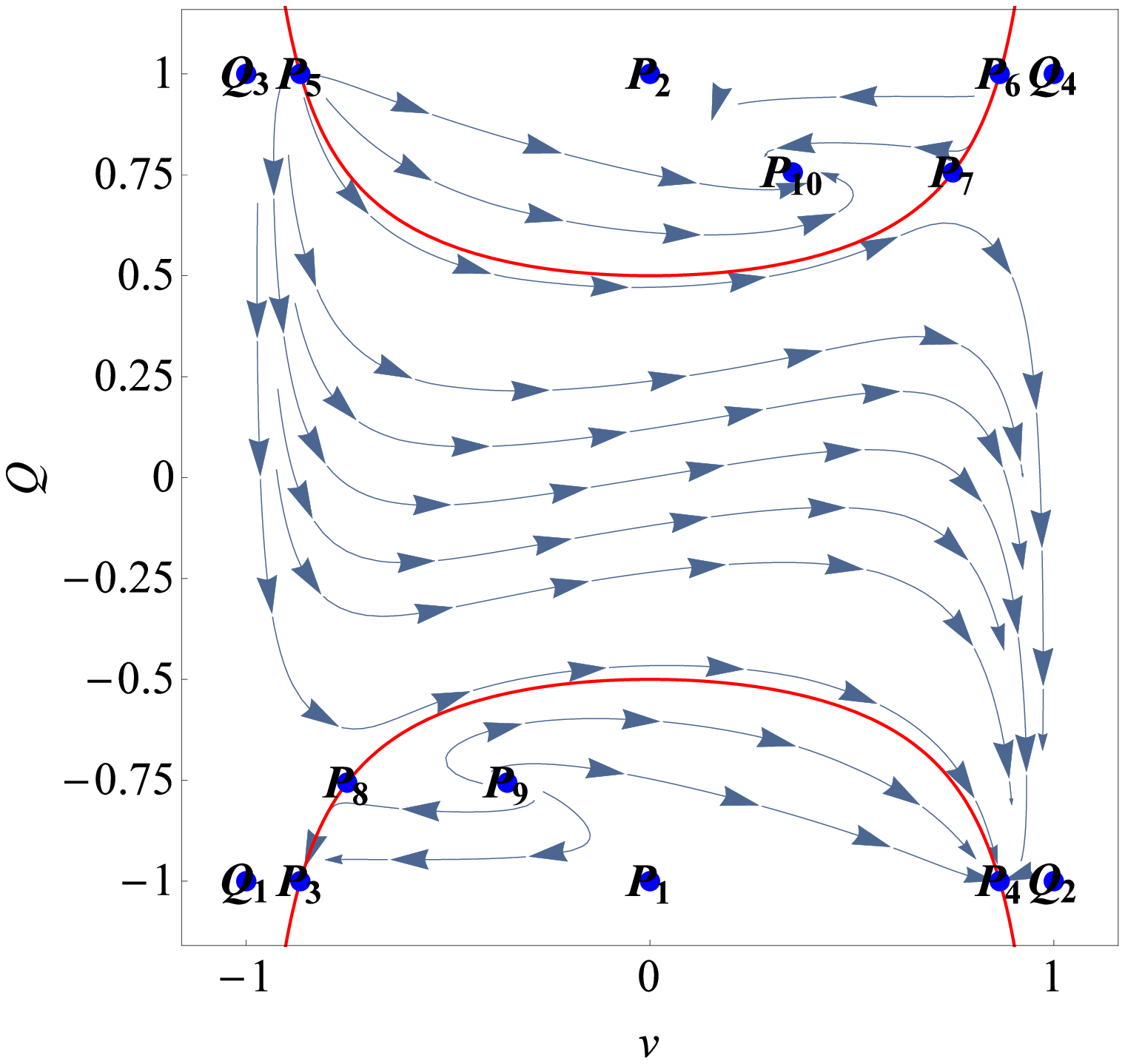}}
\subfigure[$c_1=\frac{1}{2}, c_3=\frac{1}{4}, \gamma=1$]{\includegraphics[width=0.4\textwidth]{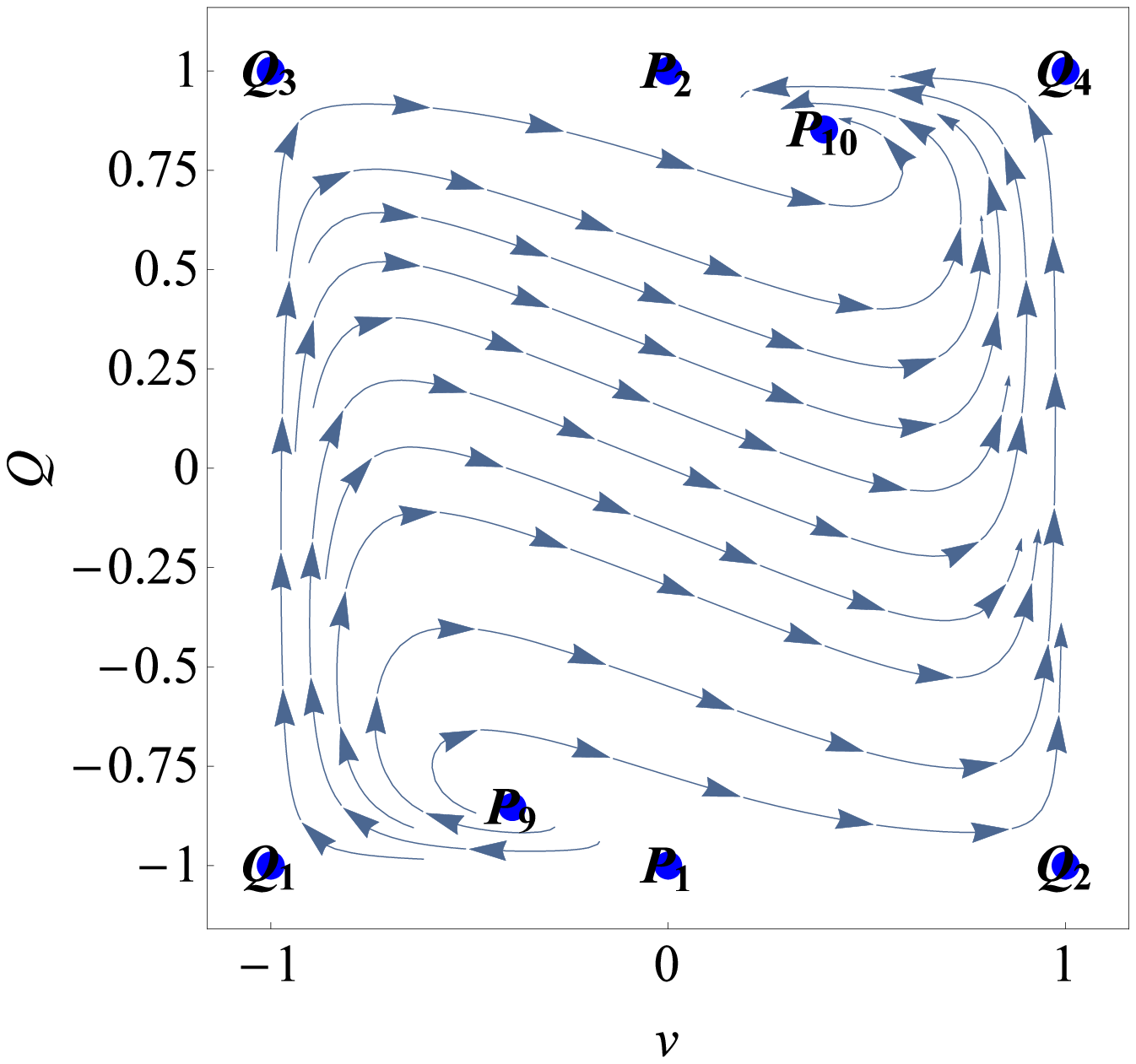}}
\caption{Streamline plot of the vector field  \eqref{example2-compact}. Red (continuous) lines represent the boundary of the phase space.} \label{fig:B2-b} \end{figure}

We choose $c_a = -\frac{(c_1{^2} +  c_3{^2})}{c_1} \leq 0$, $0 \leq c_\sigma
= {c_1}
+ {c_3} \leq 1$ and $c_\theta = -\frac{(c_1{^2} -  c_3{^2})}{3c_1} \leq 0$ which means $C_1=2(c_1+c_3)-1, C_2=1-c_1+\frac{c_3^2}{c_1}$. \footnote{Do not confuse the $c_i$'s in this section with capital $C_i$'s in the previous analysis.}
In this special case the system \eqref{KS_gen_syst} becomes
{\small
		\begin{subequations}
		\label{example2}
		\begin{align}
		& y'=
y \left(\frac{Q^3
   \left(c_3^2-c_1^2\right)}{\sqrt{3}
\left((c_1-1)c_1-c_3^2\right)}+\frac{Q\left(-\frac{c_1}{-c_
1^2+c_1+c_3^2}-2\right)}{\sqrt{3}}\right)+\frac{\sqrt{3} c_1 Q y^3 (2
c_1+2c_3-1)}{(c_1-1)c_1-c_3^2}
 \nonumber\\ & +\frac{Q^2}{\sqrt{3} (-2 c_1-2
   c_3+1)}+\frac{1}{\sqrt{3} (2 c_1+2
c_3-1)}+\left(\frac{1}{\sqrt{3}}-\frac{Q^2}{\sqrt{3}}\right)
y^2,
\\
	& Q'= \frac{Q^4  \left(c_3^2-c_1^2\right)}{\sqrt{3}
\left((c_1-1)c_1-c_3^2\right)}+\frac{Q^2\left(
1-\frac{2
c_1}{-c_1^2+c_1+c_3^2}\right)}{\sqrt{3}}+\frac{c_1}{\sqrt{3}\left(-c_1^2+c_1+c_3^2\right)}
\nonumber \\
	& +y^2 \left(\frac{\sqrt{3}   c_1 Q^2 (2 c_1+2 c_3-1)}{(c_1-1)
c_1-c_3^2}+\frac{\sqrt{3}c_1 (-2 c_
1-2c_3+1)}{(c_1-1)c_1-c_3^2}\right)+\left(\frac{Q}{\sqrt{3}}-\frac{Q^3}{\sqrt{3}}\right)
y.
\end{align}
		\end{subequations}
	}	
			The phase space becomes
				\begin{equation}
		\left\{(y,Q): Q^2 \left(c_1-\frac{c_3^2}{c_1}\right)+y^2 (1-2
   (c_1+c_3))\leq 1, -1\leq Q \leq 1\right\}.
		\end{equation}
It is compact for
\begin{enumerate}
  \item $c_1<0, -c_1\leq c_3\leq \frac{1}{2} (1-2
   c_1)$ (see figure \ref{fig:B} (a)) or
	\item $0<c_1\leq \frac{1}{4}, -c_1\leq
   c_3\leq c_1$ (see figure \ref{fig:B} (b)) or
	\item $c_1>\frac{1}{4},
   -c_1\leq c_3\leq \frac{1}{2} (1-2 c_1)$ (the flow is topologically equivalent to the previous case).
\end{enumerate}
It is unbounded for
\begin{enumerate}
 \item $c_1<0, \frac{1}{2} (1-2 c_1)<c_3\leq
   1-c_1$ (see figure \ref{fig:B2-b} (a)) or
	\item $\frac{1}{4}<c_1\leq \frac{1}{2},
   \frac{1}{2} (1-2 c_1)<c_3\leq c_1$  (see figure \ref{fig:B2-b} (b)) or
	\item $c_1>\frac{1}{2}, \frac{1}{2} (1-2 c_1)<c_3\leq
   1-c_1$  (the flow is topologically equivalent to the previous
case).
\end{enumerate}
The conditions $0\leq \gamma <2, c_{\theta }<-\frac{1}{3}, c_\sigma<\frac{1}{2}$  for
which $P_7$
is a late-time accelerated solution and $P_8$ is an early-time accelerated, inflationary
solution,
now become
\begin{enumerate}
\item $c_1<0, c_3<-\sqrt{(c_1-1) c_1}$ or
\item $c_1<0, \sqrt{(c_1-1) c_1}<c_3<\frac{1}{2} (1-2 c_1)$.
\end{enumerate}
In the case of non-compact phase space we use the system \eqref{infinity} which becomes
{\small
\begin{subequations}
\label{example2-compact}
{\small\begin{align}
&v'=\frac{Q v^5 \left(c_1^2 \left(6 \gamma +(3 \gamma -2) Q^2-16\right)+6 (\gamma -2) c_1
   (c_3-1)+c_3^2 \left((2-3 \gamma ) Q^2+4\right)\right)}{2 \sqrt{3} \left((c_1-1)
   c_1-c_3^2\right)} \nonumber \\ & -\frac{Q v^3 \left(c_1^2 \left(6 \gamma +(6 \gamma -4)
Q^2-20\right)+3
   (\gamma -2) c_1 (2 c_3-3)+2 c_3^2 \left((2-3 \gamma ) Q^2+4\right)\right)}{2 \sqrt{3}
   \left((c_1-1) c_1-c_3^2\right)} \nonumber \\ & +\frac{Q v \left(c_1^2 \left((3 \gamma
-2)
   Q^2-4\right)-3 (\gamma -2) c_1+c_3^2 \left((2-3 \gamma ) Q^2+4\right)\right)}{2
\sqrt{3}
   \left((c_1-1) c_1-c_3^2\right)} \nonumber \\ & -\frac{\left(Q^2-1\right) \sqrt{3-3 v^2}
v^2 (2 c_
1+2
   c_3-3)}{6 c_1+6 c_3-3} \nonumber \\ & -\frac{\left(Q^2-1\right) \sqrt{3-3 v^2}}{6 c_1+6
   c_3-3}+\frac{2 \left(Q^2-1\right) \sqrt{3-3 v^2} v^4 (c_1+c_3-1)}{6 c_1+6
   c_3-3},
	\\
& Q'=	-\frac{\left(Q^2-1\right) v^2 \left(c_1^2 \left(6 (\gamma -2)+(3 \gamma -2)
Q^2\right)+2
   c_1 (-3 \gamma +3 (\gamma -2) c_3+4)+(2-3 \gamma ) c_3^2 Q^2\right)}{2 \sqrt{3}
   \left((c_1-1) c_1-c_3^2\right)} \nonumber \\ & +\frac{(3 \gamma -2) \left(Q^2-1\right)
\left(Q^2
   (c_1-c_3) (c_1+c_3)-c_1\right)}{2 \sqrt{3} \left((c_1-1)
   c_1-c_3^2\right)}-\frac{Q \left(Q^2-1\right) \sqrt{1-v^2} v}{\sqrt{3}}.
	\end{align}}
\end{subequations}
}
In figures \ref{fig:B} are presented some phase plots of the system \eqref{example2}.
There, the
late-time attractors are the stiff-like solutions ($q=2$) $P_3$ and/or $P_4$.  The transition from an expanding to a contracting universe is demonstrated numerically (e.g., attractor $P_3$,
sources $P_5$ \&
 $P_6$).

In figure \ref{fig:B2-b} are presented some phase plots of the system
\eqref{example2-compact}. In
Fig. \ref{fig:B2-b} (a), the sinks are $P_{10}$ with eigenvalues $\{-0.286411+0.252591
i,-0.286411-
0.252591 i\}$ (it is a stable focus) and $P_4$ with eigenvalues $\{-0.433013,-0.0773503\}$
(it is a
stable node). The physical portion of the phase space is enclosed by the red lines
(hyperbolaes).
Hence, the only physical sink for the choice $c_1=-\frac{1}{2}, c_3=\frac{3}{2}, \gamma=1$
is $P_4$.
 In figure \ref{fig:B2-b} (b) we use the values $c_1=\frac{1}{2}, c_3=\frac{1}{4},
\gamma=1$. The
sink is $P_{10}$ with eigenvalues $\{-0.312463+0.609103 i,-0.312463-0.609103 i\}$. The
critical points
$P_3$-$P_8$ do not exist (since for them $Q>1$).
In the plots the points $Q_1$-$Q_4$ are saddles.
They
represent the points at infinity of the system \eqref{example2}. Additionally, in Figs.
\ref{fig:B2-b} (b,c), it is clearly illustrated that the transition from the decelerated
contracting solution $P_9$ to the decelerated expanding solution $P_{10}$ is in fact valid.

 \subsubsection{Case B(ii)}
	Setting $c_1=c_3$ in case B, and redefining $c_1=\frac{1}{4}(1-c^2)$, we obtain $c_\sigma=\frac{1}{2}(1- c^2),  c_a=-\frac{1}{2}(1-c^2), c_\theta=0$. The condition $0\leq c_\sigma \leq 1$ implies $-1\leq c\leq 1$. Without the loss of generality
we can choose $c > 0$. Since the Lagrangian is independent of $c_a$ for the Kantowski-Sachs
metric, the same results follow as for case A discussed in subsection \ref{CaseA}.
		
 \subsection{Case C}
	
	Substituting the values $c_\sigma=\frac{1}{2}(1- c^2)  \geq 0,
 		c_\theta = -\frac{1}{3}(1- c^2) \leq 0$ and rescaling the equations by the
factor $2\sqrt{3}
c^2$ (i.e., using the time reescaling $t\rightarrow t/(2\sqrt{3} c^2), c\neq 0$) we
obtain
				\begin{subequations}
		\label{system_C}
		\begin{align}
		 & y'=\left(c^2-1\right) (3 \gamma -2)
   Q^3 y+2 Q^2 \left(1- c^2 y^2\right) \nonumber \\ & +Q
   y \left(c^2 \left(-3 (\gamma -2)
   y^2-4\right)+3 \gamma -2\right)+2
   c^2 y^2-2,\\
		 & Q'=\left(Q^2-1\right)
   \left(\left(c^2-1\right) (3 \gamma
   -2) Q^2-2 c^2 Q y-3 (\gamma -2) c^2
   y^2+3 \gamma -2\right),
		\end{align}
		\end{subequations}
		defined on the phase space $$\{(y,Q): \left(1-c^2\right) Q^2+c^2 y^2\leq
1, -1\leq Q\leq 1, c^2\leq 1\}.$$
			
\begin{figure}[htb] \centering
\subfigure[$c=\sqrt{\frac{3}{5}}$ and $\gamma=1$]{\includegraphics[width=0.45\textwidth]{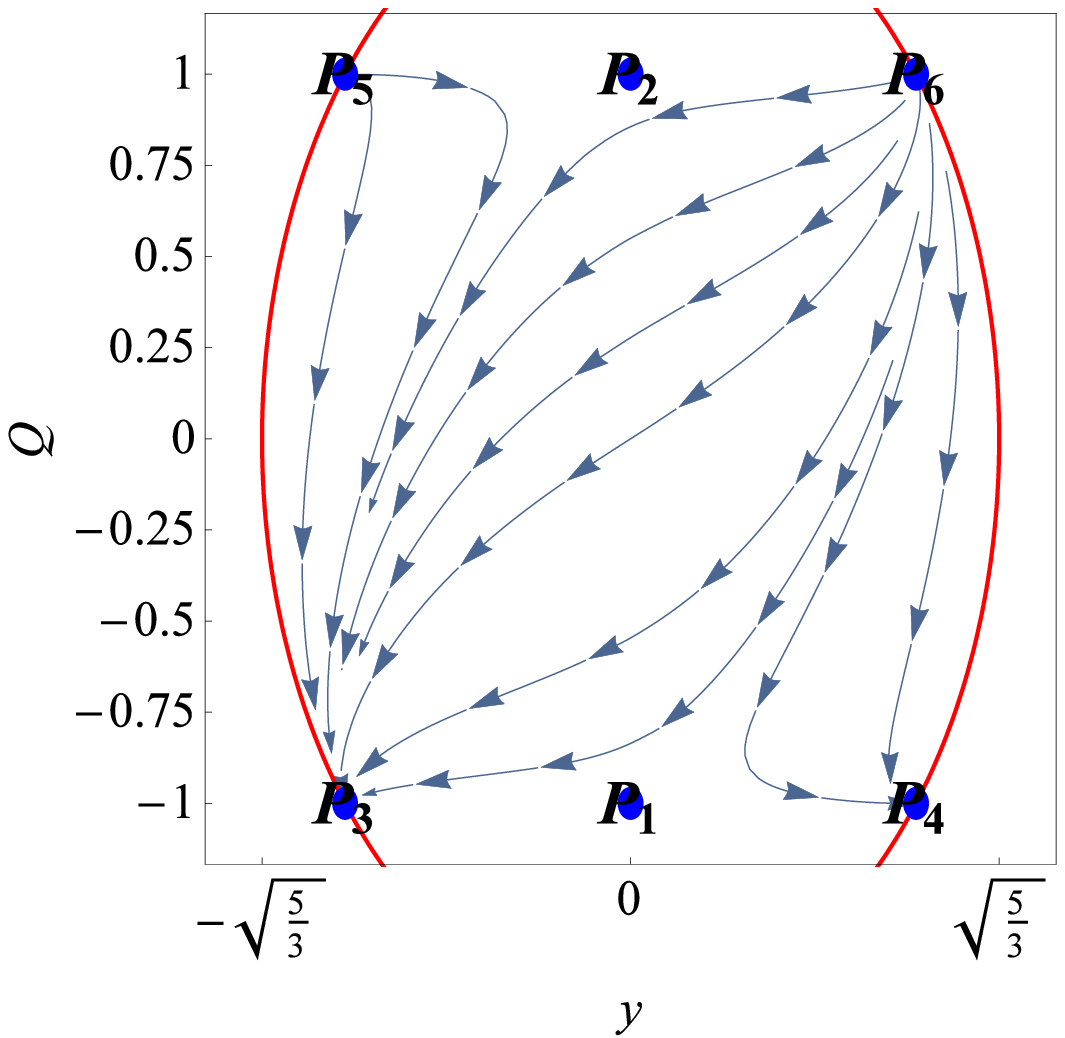}}
\subfigure[$c=0.3$ and $\gamma=1$.]{\includegraphics[width=0.45\textwidth]{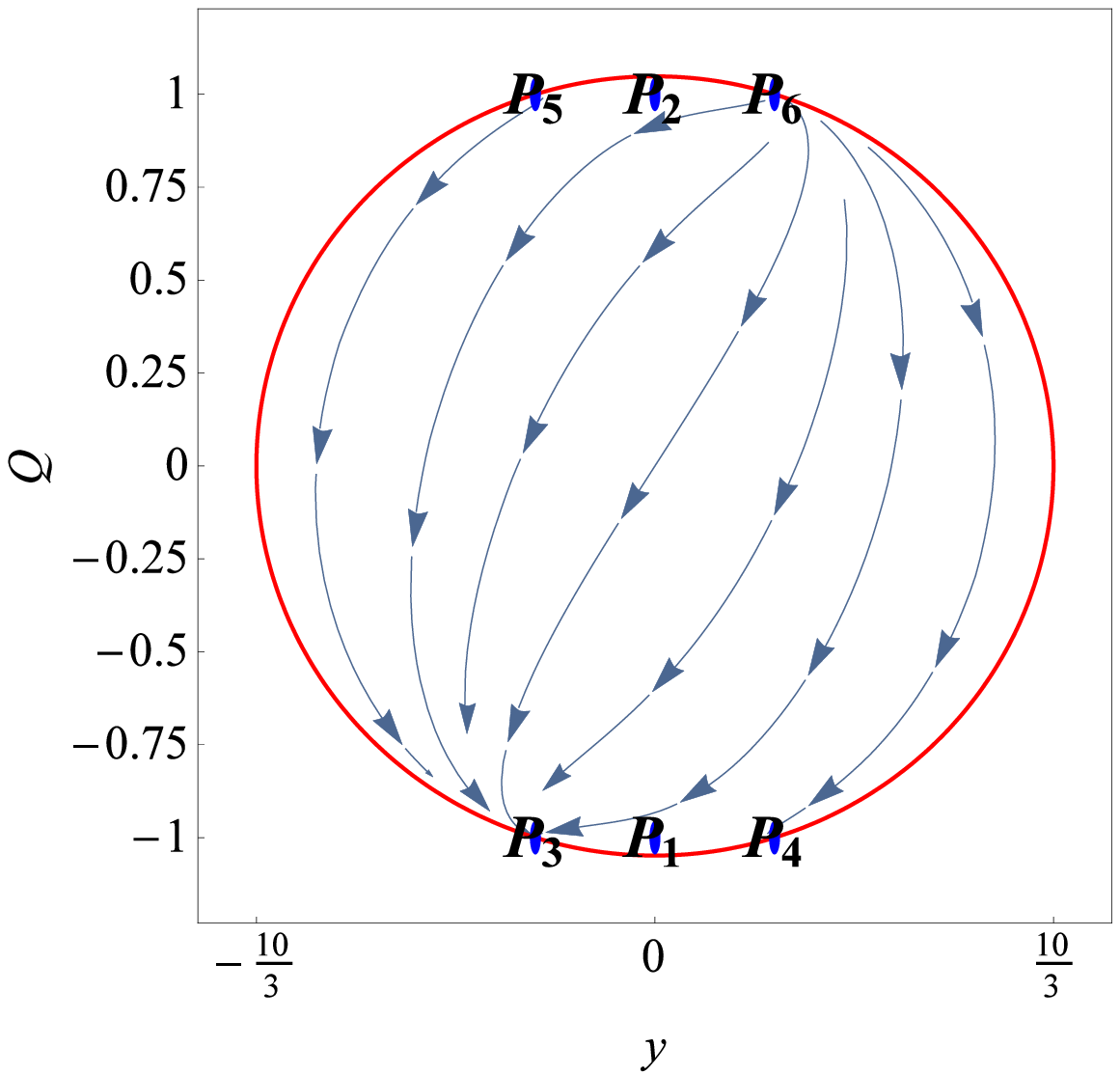}}
\subfigure[$c=0.3$ and $\gamma=0$. ]{\includegraphics[width=0.45\textwidth]{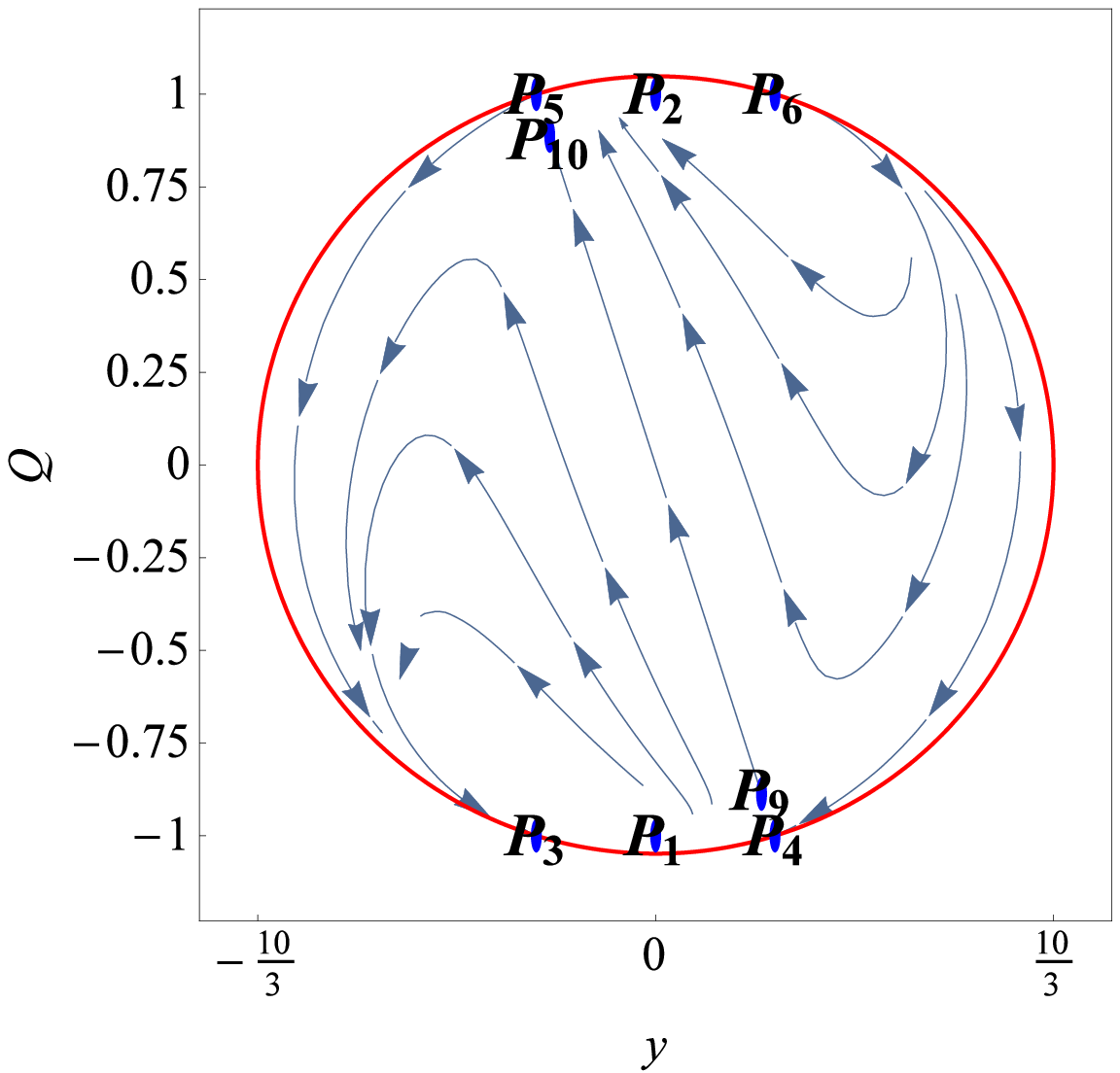}}
\subfigure[$c=0.6$ and $\gamma=0$. ]{\includegraphics[width=0.45\textwidth]{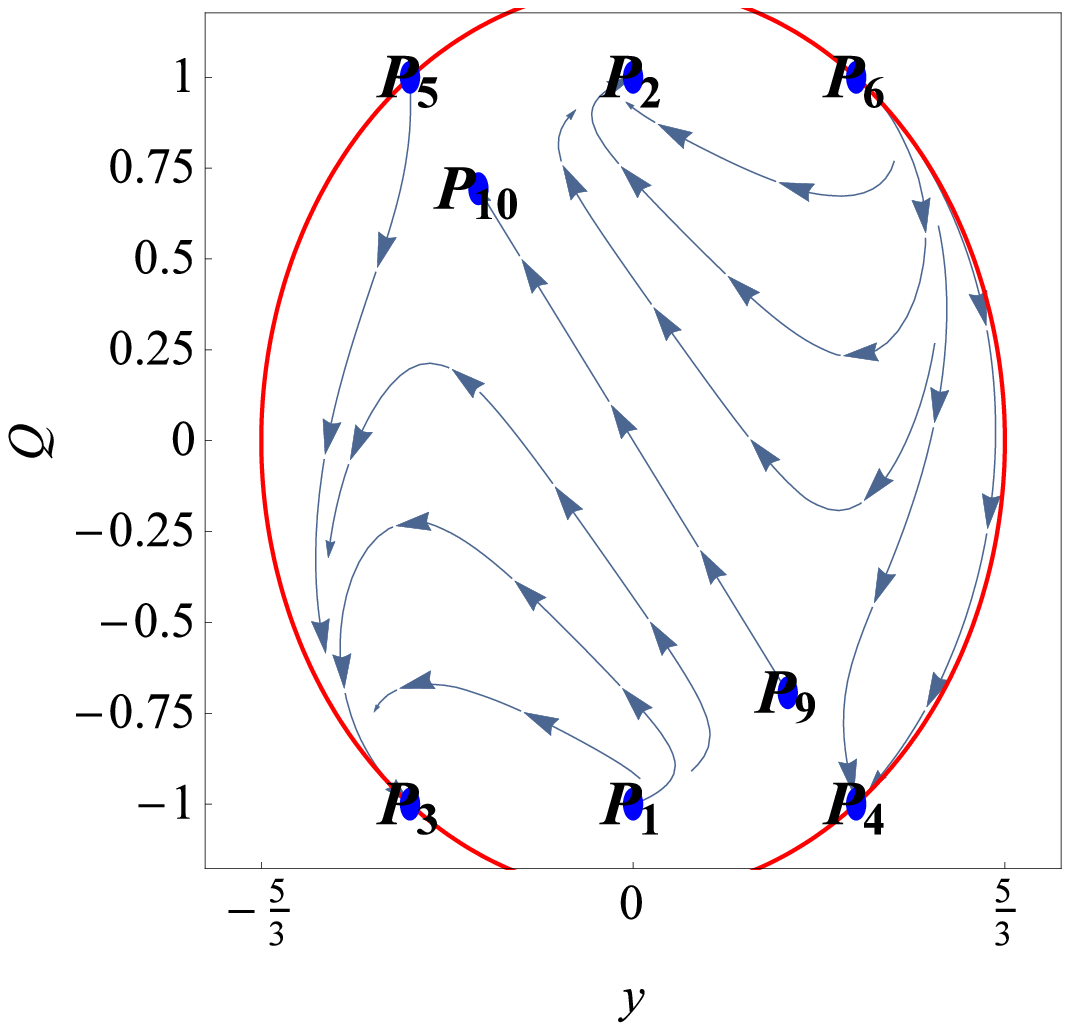}}
\caption{Streamline plot of the vector field \eqref{system_C}. Red (continuous) lines
represent
the boundary of the phase space.} \label{fig:2} \end{figure}

In this example the points $P_7$ and $P_8$	do not exist, i.e., there are no
accelerated solutions.
In the figures \ref{fig:2} (a,b) the late-time attractors are the stiff-like solutions
($q=2$) $P_
3$ and/or $P_4$. Thus, this is a clear illustration that there is a transition from an
expanding to a
contracting universe (e.g., attractor $P_3$, source $P_5$ \& $P_6$).
Additionally, we have presented some numerics for the special case
$\gamma=0$ which corresponds to a Cosmological Constant due to its cosmological interest. In this
case the anisotropic solutions $P_9$ and $P_{10}$ become saddles. As shown in Figures \ref{fig:2}
(c,d), we have the sink $P_2$, which corresponds to an accelerated de Sitter solution ($q=-1$).
Additionally, we have solutions starting at decelerated isotropic de Sitter solutions like $P_1$ and
ending up with decelerated  anisotropic solutions like $P_3$. Furthermore, we have solutions
starting with the expanding decelerated solution $P_6$, becoming a decelerated contracting solution near
the anisotropic solution $P_9$, and ending up at the expanding de Sitter solution $P_2$.

 \section{Discussion \& outlook}

In this paper we have studied Kantowski-Sachs Einstein-\ae ther perfect fluid models using
the 1+3 frame formalism \cite{WE,EU,SSSS} in the so-called comoving \ae ther gauge. The formalism
is particularly well-suited for numerical and qualitative analysis.

A special closed-form solution was found when for the perfect fluid \ the equation
of state parameter is $\frac{p}{\mu }=\gamma -1$. That special solution is
related with the existence of a group invariant transformation (Lie symmetry) for the dynamical system.
Furthermore, that special solution gives us the dominant behavior of the
system close to the movable singularity of the differential equation and by applying the ARS algorithm we found that the field equations form an integrable system. Specifically we
showed that the resonances which provides the Laurent expansion are always positive, for $\gamma \in \lbrack 0,2)$,
which means that the solution is expressed as a Right Painlev\'{e}
Series.  This means that in the complex plane the integration evolves from the singularity until a border (for details see \cite{Andriopoulos}).

In studying our model, it became apparent the system was not necessarily bounded unless
$1-2c_\sigma \geq 0$.  In the case where the phase space is bounded, we found an inflationary source at
early times, and an
inflationary sink at late times for the range of parameters $0\leq \gamma < 2$, $c_\theta
<  -\frac{
1}{3}$ and $c_\sigma < \frac{1}{2}$.
For non-compact phase space, we were able to analyze the system at infinity
by introducing a compactification scheme, and recasting the system in new variables.

We have presented three applications:
\begin{itemize}
\item  Case A: $c_\sigma=\frac{1}{2}(1- c^2)  \geq 0,
 		c_a =  -\frac{d}{(1+d)} c_\sigma \leq 0, c_\theta=0$.
		
		Under the rescaling $y\rightarrow y/c, t\rightarrow t/(2\sqrt{3} c), c>0$ we recover the system (5.27) investigated in
\cite{Coley:2015qqa}. We have re-obtained the results: for $c>0$, and when $\gamma
< \frac{2}{3}$, $P_2$ is the unique shear-free, zero curvature (FLRW) inflationary future attractor, and
for $0<c<\frac{1}{2}$ and $0\leq\gamma<2$ the sources and sinks are, respectively, $P_5$ \& $P_8$ and
$P_4$ \& $P_7$ (as confirmed in figure \ref{fig:1} (a)). All of these sources and sinks are anisotropic and all, except $P_7$, have zero curvature;
the sink  $P_7$ does not have zero curvature. For  $c>\frac{1}{2}$ the points $P_7$ \& $P_8$ do not exist, and the sources and sinks with non-zero shear are
$P_5$ and $P_4$, respectively, as confirmed numerically.

\item Case B: $c_a = -\frac{(c_1{^2} +  c_3{^2})}{c_1} \leq 0, 0 \leq c_\sigma  = {c_1} + {c_3} \leq 1,
 c_\theta = -\frac{(c_1{^2} -  c_3{^2})}{3c_1} \leq 0$.

For the choices
\begin{enumerate}
 \item $c_1<0, \frac{1}{2} (1-2 c_1)<c_3\leq
   1-c_1$  or
	\item $\frac{1}{4}<c_1\leq \frac{1}{2},
   \frac{1}{2} (1-2 c_1)<c_3\leq c_1$  or
	\item $c_1>\frac{1}{2}, \frac{1}{2} (1-2 c_1)<c_3\leq
   1-c_1$,
\end{enumerate}
we find that the phase space becomes unbounded. We demonstrate for this range of
parameters the
existence of solutions with infinite shear ($y=\pm \infty$). These solutions, with
extremely
high anisotropy, are of saddle type. 	As long as $y$ is infinite the quantity $|-3 c_\theta
Q^2+x^2|$ is infinite too, and  the restriction \eqref{constraints_KS_1a} is
satisfied. Additionally, since for $v=\pm1 $, $Q\rightarrow \pm 1$  according to the sign of
$\frac{\left(2c_{\sigma }-1\right)}{3 c_{\theta }+1}$, it thus follows that extremely high anisotropic
solutions also
have zero curvature ($K\rightarrow 0$) and infinite matter energy density ($|x|\rightarrow
\infty$)
in comparison with the Hubble scalar.
The possible sinks can be either $P_4$ or $P_{10}$, or both, in some special cases.
The attractor $P_4$ mimics a stiff-solution (i.e., $q = 2$) that is always decelerated. Thus, it is not a good
description of the late-time dynamics of the universe. Nor is $P_{10}$ a good description of the
late-time universe for $\gamma>\frac{2}{3}$, since the deceleration factor evaluated at the critical point is given by $q=\frac{3
\gamma }{2}-1$. However, the solution is always accelerated for $0\leq\gamma<\frac{2}{3}$, and then, the matter fluid behaves as dark energy. 	

\item Case C: $c_\sigma=\frac{1}{2}(1- c^2)  \geq 0,
 		c_\theta = -\frac{1}{3}(1- c^2) \leq 0, c_a = 0$.
		
		Under the time rescaling $t\rightarrow t/(2\sqrt{3} c^2), c\neq 0$, we find that the late-time attractors can be $P_3$ and $P_4$,
which mimic stiff-solutions (i.e., $q = 2$); that is, they are always decelerated and do not accurately represent
the late-time universe described by observations.
\end{itemize}
		
From the cosmological point of view, the some of the more relevant critical points are $P_7$ and $P_8$
since they can describe powerlaw accelerated solutions. Their existence conditions are:
	\begin{enumerate}
	\item $c_{\sigma }<\frac{1}{2}, c_{\theta }\leq -\frac{1}{3}$ or
	\item $c_{\sigma }<\frac{1}{2},  c_{\theta }\geq \frac{1}{3} \left(3-8 c_{\sigma
}\right)$ or
	\item $c_{\sigma }>\frac{1}{2}, \frac{1}{3} \left(3-8 c_{\sigma
   }\right)\leq c_{\theta }\leq -\frac{1}{3}$.
	\end{enumerate}
The deceleration parameter evaluated at the critical points is given by $q=-\frac{3 c_{\theta
}+1}{2 \left(2 c_{\sigma }-1\right)}$, thus the critical points are:
					\begin{enumerate}
					\item an accelerated solution for
					      \begin{enumerate}
								\item $0\leq \gamma <2,
c_{\theta }<-\frac{1}{3}, c_{\sigma }<\frac{1}{2}$ ($P_7$ is a late-time accelerated dark energy dominated solution, while $P_8$ is an  accelerated inflationary early-time solution).
								\end{enumerate}
								
          \item a decelerated solution for
                 \begin{enumerate}
								  \item $0\leq \gamma <2,
c_{\theta }<-\frac{1}{3}, c_{\sigma }\geq \frac{3}{8}
   \left(1- c_{\theta }\right)$ ($P_7$ and $P_8$ are saddles) or
								  \item $0\leq \gamma <2,
c_{\theta }>-\frac{1}{3}, \frac{3}{8} \left(1- c_{\theta
   }\right)< c_{\sigma }<\frac{1}{2}$ ($P_7$ is a sink and $P_8$ is a source).
								 \end{enumerate}
					\end{enumerate}

Furthermore, for $P_{7,8}$ we find that for $c_{\theta }<-\frac{1}{3}, \frac{3}{4}\left(c_{\theta }+1\right)<c_{\sigma}<\frac{1}{2}$, the universe does not have a Big Bang or a Big Crunch solution. In such case the point $P_7$ is always a sink and $P_8$ is always a source.

Finally, these solutions do not isotropize at late times. In fact, the criterion of late-time isotropization in an expanding universe ($\theta>0$) is the vanishing
of the shear $\sigma$ \cite{Mendes}, or alternatively we can use the stronger condition $\sigma/\theta\rightarrow 0$ as $t\rightarrow +\infty$ \cite{isotropization}. However, for $P_7$ we have, from \eqref{expressions_5.7}, $\frac{\sigma}{\theta}=\frac{1}{3} \left(\frac{1}{p}-1\right)+\mathcal{O} (\Delta t)^{1-3 p}$ which tends to $-\frac{3 c_\theta +1}{6 (2 c_\sigma -1)}\neq 0$ as  $t\rightarrow +\infty$ if $c_{\theta }< -\frac{1}{3}, c_{\sigma }<\frac{3}{4} \left(c_{\theta }+1\right)$ or $c_{\theta }<-\frac{1}{3}, c_{\sigma }>\frac{3}{8} \left(1-c_{\theta }\right)$. While for the choices $c_{\theta }<-\frac{1}{3}, \frac{3}{4} \left(c_{\theta }+1\right)<c_{\sigma }<\frac{1}{2}$ or $c_{\theta }>-\frac{1}{3}, \frac{3}{8} \left(1-c_{\theta }\right)<c_{\sigma }<\frac{1}{2}$, the term $\mathcal{O} (\Delta t)^{1-3 p}$ becomes infinite and again the model does not isotropize at late times. The same result is valid for $P_8$ after a time reversal. For other anisotropic inflationary models see \cite{anis1,anis2,anis3} and references therein.

Summarizing, it is well known that Kantowski-Sachs models in GR have two asymptotic scenarios: (i) all models
expand from
a singularity, reach a point of maximum expansion, and then recollapse to a singularity;
and (ii),
there are solutions that expand from singularities to infinitely dispersed isotropic
states and
solutions that contract from infinitely dispersed isotropic states to singularities
\cite{Coley:2003mj,Heinzle:2004sr}.  Now, in our scenario, i.e. a perfect fluid in
Kantowski-Sachs \AE-theory without scalar field, in addition to the two
asymptotic scenarios  (i) and (ii) mentioned above, we also found solutions
that either expand from or contract to anisotropic states which not be in accordance with a Big Bang or a Big Crunch behavior as mentioned above. This result, up to
our knowledge, is new (a partial proof of this was first given in \cite{Coley:2015qqa}) and do not arise
in GR. They are a non-trivial consequence of the presence of a non-zero
Lorentz-violating vector field.
We are now
exploring how to get accelerated, power-law isotropic solutions in Kantowski-Sachs
\AE-theories.
Particularly, by including an additional scalar field with a self-interaction potential
which
depends not only on the scalar field, but also on the shear and the expansion parameter of
the \ae ther. Einstein-\ae ther  models with an exponential potential were recently studied
in \cite{Barrow:2012qy,Sandin:2012gq,Alhulaimi:2013sha}. We expect to make further progress
on this question.

\noindent
\section*{\em Acknowledgements}

This work was supported, in part, by NSERC of Canada.
G.L. was supported  by FONDECYT grant no. 3140244.
AP acknowledges financial support of FONDECYT grant no. 3160121.
G.L thanks his family and beautiful twins for emotional support
during the preparation of this work. The obstetric team of Hospital Carlos Van Buren,
and neonatology of both Hospital Carlos Van Buren and Cl\'{\i}nica Ciudad del Mar, are
acknowledged for postnatal care.
Alan Coley, P.G.L. Leach and E. N. Saridakis are acknowledged for helpful discussions. G.L.  also thanks National Technical University of Athens, and to Universidad Austral de Chile, for warm hospitality during the final stages of this research.
AP thanks the University of Athens for the hospitality while part of this work carried out.


\end{document}